%% file: ms.tex
\documentclass[
aip,jcp,amsmath,amssymb,
preprint,tightenlines,longbibliography]{revtex4-1}
\pdfoutput=1

\usepackage{bm}

\usepackage{graphicx}

\newcommand{\mean}[1]{\ensuremath{\left<#1\right>}}

\newcommand{\df}{\ensuremath{\mathrm{d}}}

\newcommand{\va}{\ensuremath{\mathbf{a}}}
\newcommand{\vmu}{\ensuremath{\bm{\mu}}}
\newcommand{\vnu}{\ensuremath{\bm{\nu}}}
\newcommand{\vth}{\ensuremath{\bm{\theta}}}
\newcommand{\vphi}{\ensuremath{\bm{\phi}}}
\newcommand{\vlam}{\ensuremath{\bm{\lambda}}}
\newcommand{\vx}{\ensuremath{\mathbf{x}}}
\newcommand{\hatvth}{\ensuremath{\bm{\hat{\theta}}}}
\newcommand{\vzero}{\ensuremath{\mathbf{0}}}

\newcommand{\pE}{\ensuremath{\mathbb{E}}}

\newcommand{\opL}{\ensuremath{\mathcal{L}}}

\newcommand{\natz}{\ensuremath{\mathbb{N}_0}}
\newcommand{\pos}{\ensuremath{\mathbb{R}_{\geq 0}}}

\begin{document}

\preprint{The following article has been submitted to the Journal of Chemical Physics.}

\title{A variational approach to moment-closure approximations for the kinetics of biomolecular reaction networks} 

\affiliation{Department of Electrical Engineering and Information Technology, Technische Universit\"at Darmstadt,\\64283 Darmstadt, Germany}

\author{Leo Bronstein}

\author{Heinz Koeppl}
\email[]{heinz.koeppl@bcs.tu-darmstadt.de}

\date{\today}

\begin{abstract}
Approximate solutions of the chemical master equation and the chemical Fokker-Planck equation are an important tool in the analysis of biomolecular reaction networks.
Previous studies have highlighted a number of problems with the moment-closure approach used to obtain such approximations, calling it an ad-hoc method.
In this article, we give a new variational derivation of moment-closure equations which provides us with an intuitive understanding of their properties and failure modes and allows us to correct some of these problems.
We use mixtures of product-Poisson distributions to obtain a flexible parametric family which solves the commonly observed problem of divergences at low system sizes.
We also extend the recently introduced entropic matching approach to arbitrary ansatz distributions and Markov processes, demonstrating that it is a special case of variational moment closure. This provides us with a particularly principled approximation method.
Finally, we extend the above approaches to cover the approximation of multi-time joint distributions, resulting in a viable alternative to process-level approximations which are often intractable.
\end{abstract}


\maketitle 

\input{intro}

\input{theory}

\input{examples}

\input{analysis}

\input{further_applications}

\input{conclusion}

\begin{acknowledgments}
H.K. acknowledges support from the LOEWE research priority program CompuGene.
\end{acknowledgments}

\input{appendix_gaussian}

\bibliography{references}

\end{document}

%% file: intro.tex
\section{Introduction}
The solution of the chemical master equation (CME) or of the chemical Fokker-Planck equation (CFPE) for a chemical reaction network can typically not be obtained in closed form, and numerical approaches are computationally infeasible for larger systems.
One established technique for obtaining approximate solutions is the linear noise approximation \cite{vankampen1983}, which however assumes sufficiently large system sizes.
In biological systems, where the abundance of certain chemical species can be very low, the linear noise approximation might become inappropriate.
The main alternative is the use of moment equations in combination with moment closure, a method also employed in many other fields \cite{kuehn2016}.
Moment closure approximations have often been referred to as ad-hoc approximations, and it has been shown \cite{schnoerr2014, schnoerr2015} that they can exhibit unphysical behavior.
Another approximation which has recently been proposed and relies on information-theoretic considerations is entropic matching \cite{ramalho2013}. It employs a Gaussian approximating distribution.

In this article, we unify and extend the above approaches.
We provide a variational derivation of moment closure which exhibits it as a principled approximation and helps us to understand some of the failure modes typically observed.
We also extend entropic matching to general Markov processes and arbitrary approximating distributions and show it to be a special case of variational moment closure.
While variational moment closure is a principled approximation, only a subset of all possible closure schemes can be justified by it.
On the one hand, we demonstrate that this subset does not suffer from some of the problems often attributed to ad-hoc closure schemes.
For this purpose, we introduce mixtures of independent Poisson distributions as a general and useful class of closure distributions.
On the other hand, some of the problems attributed to moment closure are not resolved by the variational approach (and are also present in entropic matching).
However, our new variational interpretation of moment closure does provide an intuitive explanation for these failure modes.
While we present our results in the context of chemical kinetics, they are valid more generally for approximations of other Markov processes.

We begin by deriving and connecting variational moment closure and entropic matching in Section~\ref{sec:theory}. We demonstrate these methods on several examples in Section~\ref{sec:entmatch}. In Section~\ref{sec:poisson_closures} we analyze the deficiencies of moment closure, where we also suggest possible solutions.
Finally, in Section~\ref{sec:further_app} we generalize variational moment closure and entropic matching to the approximation of multi-time joint distributions.

%% file: theory.tex
\section{Variational moment closure and entropic matching} \label{sec:theory}
Throughout, we consider a reaction network consisting of $N$ species and $R$ reactions
\begin{equation} \label{eq:reaction_system}
\sum_{n=1}^N{s_{nj} \mathcal{X}_n} \longrightarrow \sum_{n=1}^N{r_{nj} \mathcal{X}_n}, \quad j = 1, \dots, R.
\end{equation}
The established stochastic process model for such a reaction network, taking into account the discreteness of molecule counts, is a continuous-time Markov chain.
The marginal probabilities $p_t(\vx) = P(\mathbf{X}_t = \vx)$ of the process $\mathbf{X}_t \in \natz^N$ describing molecule counts are then governed by the CME
\begin{equation} \label{eq:cme}
\frac{\partial p_t(\vx)}{\partial t} = \sum_{j=1}^R\left\{ h_j(\vx-\vnu_j) p_t(\vx-\vnu_j) - h_j(\vx) p_t(\vx)\right\}.
\end{equation}
Here $\vnu_j = (\nu_{1j}, \dots, \nu_{Nj}) = (r_{1j}-s_{1j}, \dots, r_{Nj}-s_{Nj})$ is the stoichiometric change vector and $h_j(\vx)$ the reaction hazard of the $j$-th reaction when the system is in state $\vx = (x_1, \dots, x_N)$. 
In all concrete examples that we consider, we assume mass-action kinetics, so that the $h_j$ are given by
\begin{equation} \label{eq:mass_action}
h_j(\vx) = h_j(\vx, \Omega) = \Omega c_j \prod_{n=1}^N{\frac{(x_n)_{s_{nj}}}{\Omega^{s_{nj}}}},
\end{equation}
where $(x)_s = x(x-1)\cdots(x-s+1)$ denotes the falling factorial. Here we have introduced the system size $\Omega$ in terms of which it will be convenient to analyze the behavior of the approximation schemes discussed in this article.
For some purposes, it is sufficient to consider an approximation based on a continuous description in terms of concentrations $\vx \in \pos^N$, so that the underlying stochastic process model becomes a stochastic differential equation.
The marginal probability density is then governed by the Fokker-Planck equation
\begin{equation} \label{eq:fpe}
\frac{\partial p_t(\vx)}{\partial t} = -\sum_{n=1}^N{\frac{\partial [a_n(\vx) p_t(\vx)]}{\partial x_n} }
+ \frac{1}{2} \sum_{n,m=1}^N{ \!\!\! \frac{\partial^2 [B_{nm}(\vx) p_t(\vx)]}{\partial x_n \partial x_m} }.
\end{equation}
The coefficients are given by\cite{gillespie2000}
\begin{equation}
\begin{aligned}
a_n(\vx) &= \sum_{j=1}^{R}{h_j(\vx) \nu_{nj}},\\
B_{nm}(\vx) &= \sum_{j=1}^{R}{h_j(\vx) \nu_{nj}\nu_{mj}}.
\end{aligned}
\end{equation}
Both CME and CFPE have the general form
\begin{equation} \label{eq:evolution}
\frac{\partial p_t(\vx)}{\partial t} = \opL p_t(\vx),
\end{equation}
where $\opL$ is the appropriate (Kolmogorov-forward) evolution operator of either the CME or the CFPE.
The derivations presented in the following use this notation, and are in fact valid for other Markov processes.
The reason why we consider both CME and CFPE is not only for generality, but also because, as we shall see, any moment-closure scheme used should be adapted to the equation one is considering.

\subsection{Variational moment closure} \label{sec:variational}
We now derive the usual moment closure equations in a way which, apart from being principled, clearly shows their significance.
Our goal is to approximate the solution of \eqref{eq:evolution} at each time $t$ by a member of a parametric family of probability distributions $p_{\vth}(\vx)$.
The parameter vector $\vth = (\theta_1, \dots, \theta_L)$ ranges in some open subset of $\mathbb{R}^L$.
Since the solution of \eqref{eq:evolution} depends on the time $t$, a full approximate solution is given by a curve $\vth(t)$.
Thus, the time-dependence in the solution $p_t(\vx)$ of \eqref{eq:evolution} is contained in the time-dependence of the parameters $\vth(t)$.
Of course, the exact solution will in general not be a member of the chosen parametric family, so we require some means of measuring the approximation error in order to define in which sense the approximation is to be performed.

To obtain the moment-closure equations, this is done in the following way:
Begin by choosing a collection of moment functions $\vphi(\vx) = (\phi_1(\vx), \dots, \phi_K(\vx))$.
For example, to obtain the usual moment equations of second order, one would choose $K = N + N(N+1)/2$ monomial moment functions
\begin{equation} \label{eq:example_momfuncs}
\begin{aligned}
\phi_{n}(\vx) &= x_n,  &n&= 1, \dots, N,\\
\phi_{nm}(\vx) &= x_n x_m, &n,m& = 1, \dots, N, \quad n \leq m.
\end{aligned}
\end{equation}
Each of these moment functions can be used to measure the distance between two distributions $p(\vx)$ and $q(\vx)$ via the difference between their means $\mean{\phi_k(\vx)}_{p(\vx)} - \mean{\phi_k(\vx)}_{q(\vx)}$. To turn this difference into a meaningful distance measure, we map it through a function $C: \mathbb{R} \rightarrow [0, \infty)$ to arrive at
\begin{equation} \label{eq:error_functions}
E_k(p,q) = C\left(\mean{\phi_k(\vx)}_{p(\vx)} - \mean{\phi_k(\vx)}_{q(\vx)}\right).
\end{equation}
As will be seen below, the precise form of this function is not relevant, as long as it satisfies $C(0) = C'(0) = 0$ and $C''(0) > 0$.
For simplicity, we will show the derivation for $C(x) = x^2/2$.

We will derive an ordinary differential equation for the parameter vector $\vth(t)$.
To do this, assume that at some time-point $t$, we have an approximation $p_{\vth(t)}(\vx)$ of the solution of \eqref{eq:evolution} available.
Allowing this distribution to evolve a short time $\Delta t$ using the evolution operator $\opL$ of the Markov process, the result
\begin{equation} \label{eq:dist_evolved}
p(\vx) = p_{\vth(t)}(\vx) + \Delta t \opL p_{\vth(t)}(\vx) + \mathcal{O}(\Delta t^2)
\end{equation}
will in general no longer belong to the parametric family.
We can try to determine a new approximation $p_{\vth(t+\Delta t)}(\vx)$ from the parametric family by choosing $\vth(t+\Delta t)$ to minimize (simultaneously) the errors
\[
E_k(p_{\vth(t)} + \Delta t \opL p_{\vth(t)}, \; p_{\vth(t+\Delta t)}), \quad\quad k = 1, \dots, K.
\]
Of course, choosing such a $\vth(t+\Delta t)$ simultaneously for all error functions will in general not be possible.
However, we are in fact only interested in the limit $\Delta t \rightarrow 0$ (in which $p_{\vth(t)} + \Delta t \opL p_{\vth(t)} \rightarrow p_{\vth(t)}$) in order to obtain an ordinary differential equation for the parameters $\vth$.
Writing for brevity $\vth = \vth(t), \hat{\vth} = \vth(t+\Delta t)$, the resulting equations for $\hatvth$ are
\begin{equation} \label{eq:deriv_step_1}
\begin{aligned}
0 &= \frac{\partial E_k(p_{\vth} + \Delta t \opL p_{\vth}, \; p_{\hatvth})}{\partial \hat{\theta}_i} \\
&= \bigg[ \mean{\phi_k}_{\vth} - \mean{\phi_k}_{\hat{\vth}}   + \Delta t \mean{\opL^{\dagger}\phi_k}_{\vth} \bigg] \frac{\partial \mean{\phi_k}_{\hat{\vth}}}{\partial \hat{\theta}_i}.
\end{aligned}
\end{equation}
Here $\opL^{\dagger}$ is the adjoint of the operator $\opL$, which acts on functions $\psi$ and is given by
\begin{equation} \label{eq:cme_backward}
\opL^{\dagger} \psi(\vx) = \sum_{j=1}^R h_j(\vx)\left\{\psi(\vx+\vnu_j) - \psi(\vx)\right\}
\end{equation}
for the CME \eqref{eq:cme} and by
\begin{equation} \label{eq:fpe_backward}
\opL^{\dagger} \psi(\vx) = \sum_{n=1}^N{a_n(\vx)  \frac{\partial  \psi(\vx)}{\partial x_n}} 
+ \frac{1}{2} \sum_{n,m=1}^N{B_{nm}(\vx) \frac{\partial^2 \psi(\vx)}{\partial x_n \partial x_m} }
\end{equation}
for the CFPE \eqref{eq:fpe}.
By $\mean{\,\cdot\,}_{\vth}$ we denote an expectation taken with respect to $p_{\vth}$.

Dividing \eqref{eq:deriv_step_1} by $\Delta t$ and taking the limit, we obtain
\[
0 = 
\left[ -\sum_{l=1}^L{\frac{\partial \mean{\phi_k}_{\vth}}{\partial \theta_l} \dot{\theta}_l} 
+ \mean{\opL^{\dagger}\phi_k}_{\vth} \right] \frac{\partial \mean{\phi_k}_{\vth}}{\partial \theta_i} .
\]
We now assume that the matrix
\begin{equation} \label{eq:matrix_M}
F_{kl}(\vth) = \frac{\partial \mean{\phi_k}_{\vth}}{\partial \theta_l} = \mean{ \phi_k \frac{\partial \ln p_{\vth}}{\partial \theta_l} }_{\vth}
\end{equation}
is invertible.
In particular, the number $K$ of moment functions and the dimension $L$ of the parameter vector $\vth$ are equal and $\nabla_{\vth} \mean{\phi_k}_{\vth} \neq \mathbf{0}$ for each $k$.
We then obtain
\begin{equation} \label{eq:moment_closure}
\dot{\vth} = F(\vth)^{-1} \mean{ \opL^{\dagger} \vphi }_{\vth}
\end{equation}
which are the moment-closure equations when parameterized in terms of $\vth$, and when using the distributional ansatz $p_{\vth}$ to close the equations.
When we instead introduce the parameters $\vmu = \mean{\vphi}_{\vth}$, we have
\[
\dot{\mu}_k = \sum_{l=1}^K{\frac{\partial \mu_k}{\partial \theta_l} \dot{\theta}_l}
= \sum_{l=1}^K{\frac{\partial \mean{\phi_k}_{\vth}}{\partial \theta_l} \dot{\theta}_l} 
= \sum_{l=1}^K{F_{kl}(\vth) \dot{\theta}_l} 
\]
so that we obtain the moment-closure equations in their usual form
\begin{equation} \label{eq:moment_closure_moments}
\dot{\vmu} = \mean{ \opL^{\dagger} \vphi }_{\vmu}.
\end{equation}
Note again that we could have replaced $C(x) = x^2/2$ by any other function as long as $C(0) = C'(0) = 0$ and $C''(0) > 0$, because we only used the error functions $E_k$ in the limit $\Delta t \rightarrow 0$.
 
The moment-closure equations can be seen, as is evident from our derivation, as a `greedy' algorithm (to borrow terminology from computer science).
The parameter vector $\vth(t)$ evolves to minimize the approximation error (as measured by $\eqref{eq:error_functions}$) after an infinitesimal time-step $\df t$, irrespective of the effect this might have on the approximation quality at a later time.
This property will help us to understand some of the failure modes of moment-closure approximations in Section~\ref{sec:poisson_closures}.
We note that a different (more complicated) variational justification for moment closure has been given by Eyink \cite{eyink1996action}.

The two ingredients for a moment-closure approximation are thus a choice of parametric distribution $p_{\vth}(\vx)$ and a choice of moment functions $\vphi(\vx)$.
We stress that the space on which these are defined has to be adapted to the Markov process under investigation.
Thus, if one is considering the CME \eqref{eq:cme}, moment functions and ansatz distribution should be defined on $\natz^N$, whereas if one is considering the CFPE, they should be defined on $\pos^N$. In Section~\ref{sec:poisson_closures}, we demonstrate that at least some of the problems usually attributed to moment-closure techniques can be explained by a failure to take this into account when choosing a distributional ansatz. 
For the purpose of this article, we will call any moment-closure approach which can be seen as an instance of variational moment closure \emph{principled}.
Thus, a moment closure scheme is principled if (i) ansatz distribution and moment functions are defined on the correct state space and (ii) the number $K$ of moment functions and the number $L$ of ansatz distribution parameters are equal and the matrix $\eqref{eq:matrix_M}$ is invertible.
Any approach which cannot be justified in this way will be called not principled (or ad-hoc), although of course there might exist other justifications for it.
Note also that moment closure, when justifiable via the variational approach, provides an approximation to the full solution of the CME or CFPE. It does not merely provide lower-order moments, as is often asserted for ad-hoc moment-closure approximations.

Our derivation does not make use of the fact that $p_{\vth}(\vx)$ is a probability distribution. Thus it is possible to allow $p_{\vth}(\vx)$ to take negative values or to not sum to one. Using parametric families $p_{\vth}(\vx)$ which can become negative in some (perhaps negligible) part of the state space provides one with more flexibility when choosing an appropriate ansatz.
Also note that the derivation above remains valid when the moment functions depend on the variational parameters, i.e. $\vphi(\vx) = \vphi_{\vth}(\vx)$, which will be necessary to establish the connection to entropic matching.
In this case, the error functions \eqref{eq:error_functions} are defined using the parameter value $\vth(t)$ when deriving the evolution equation from time $t$ to $t + \Delta t$.

\subsection{Entropic matching} \label{sec:entmatch_theory}
\begin{figure*}
\includegraphics[width=1.0\textwidth]{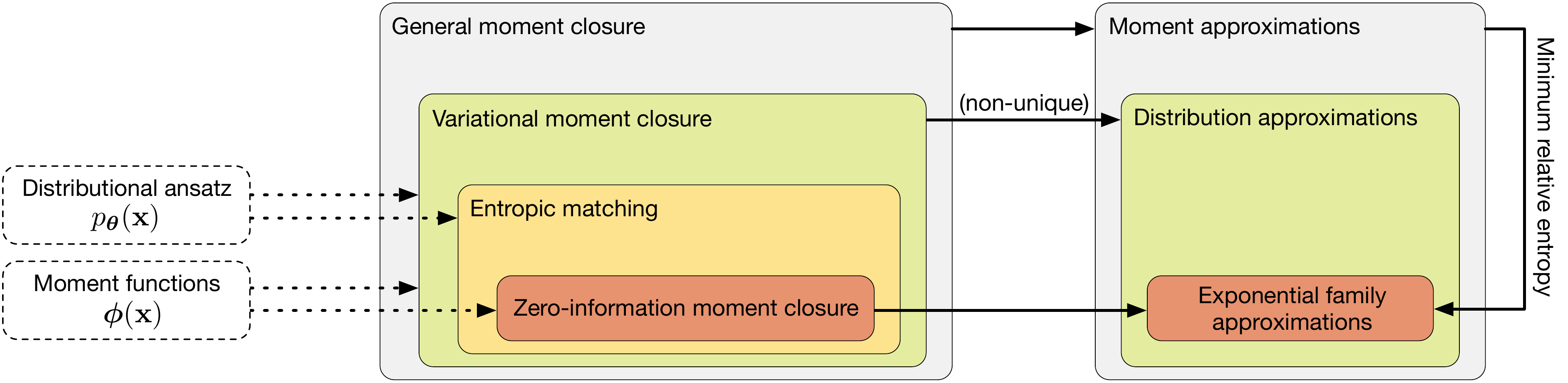}%
\caption{Relation between general (ad-hoc) moment closure, variational moment closure, entropic matching and zero-information moment closure.
Dashed arrows indicate the building blocks for each approximation:
Variational moment closures are based on a choice of moment functions $\vphi(\vx)$ and a choice of ansatz distribution $p_{\vth}(\vx)$. Entropic matching requires only a choice of ansatz distribution, while ZI moment closure requires only a choice of moment functions.
In this sense, entropic matching and ZI moment closure are dual to each other.
Full arrows indicate the type of approximation provided by each approximation method:
Ad-hoc closures provide only approximations of low-order moments, whereas variational moment closures (and the special cases entropic matching and ZI moment closure) provide approximations of the full distribution. However, ad-hoc closures can provide approximations of the full distributions via minimum relative entropy.
\label{fig:relation_diagram}}%
\end{figure*}

While the approach described in Section~\ref{sec:variational} is very flexible, it does not provide any indication of how to choose the ansatz distribution $p_{\vth}(\vx)$ and the moment functions $\vphi(\vx)$.
Here we extend entropic matching \cite{ramalho2013}, which is an approximation method based on information-theoretic considerations, to arbitrary ansatz distributions and processes. As we will see, entropic matching turns out to be a special case of variational moment closure, which provides a natural choice of moment functions $\vphi(\vx)$ for any choice of distribution $p_{\vth}(\vx)$.
The relationships between the various approximations which we derive here and in the following sections are depicted graphically in Fig.~\ref{fig:relation_diagram}.

The idea underlying entropic matching is the same as was demonstrated in our derivation of the moment-closure equations, which in fact was inspired by entropic matching.
The difference lies in the distance measure used.
From an information-theoretic point of view, there are strong arguments \cite{leike2017} for using the relative entropy 
\[
D[p \parallel q] = \mean{\ln \frac{p(\vx)}{q(\vx)}}_{p(\vx)}
\]
as a general-purpose distance measure to a distribution $p(\vx)$ when approximating it by a distribution $q(\vx)$.
We now follow the same approach as in the derivation of variational moment closure, using the relative entropy as distance measure.
We again try to find an approximate solution to \eqref{eq:evolution} within a parametric family $p_{\vth}(\vx)$.
Assume that, at some time $t$, an approximating distribution is available and specified by the parameters $\vth(t)$.
A small time $\Delta t$ later, the distribution is again given by \eqref{eq:dist_evolved}.
Entropic matching proceeds by approximating this distribution by $p_{\vth(t+\Delta t)}(\vx)$, where $\vth(t+\Delta t)$ is chosen to minimize the relative entropy $D[p \parallel p_{\vth(t+\Delta t)}]$ between $p(\vx)$ and $p_{\vth(t+\Delta t)}(\vx)$.
Writing again $\vth = \vth(t), \hatvth = \vth(t+\Delta t)$ and using \eqref{eq:dist_evolved}, the relative entropy is given, up to order one in $\Delta t$, by
\[
\begin{aligned}
&D[p \parallel p_{\hatvth}]  \\
&= \mean{\ln \frac{p_{\vth} + \Delta t \opL p_{\vth}}{p_{\hatvth}}}_{p(\vx)} \\
&= \mean{\ln \frac{p_{\vth}}{p_{\hatvth}}}_{\vth}
+ \Delta t \left[ \mean{\frac{\opL p_{\vth}}{p_{\vth}}}_{\vth}
+ \mean{\frac{\opL p_{\vth}}{p_{\vth}} \ln \frac{p_{\vth}}{p_{\hatvth}} }_{\vth} \right] \\
&= D[p_{\vth} \parallel p_{\hatvth}] + \Delta t \left[ \mean{\frac{\opL p_{\vth}}{p_{\vth}}}_{\vth}
+ \mean{\frac{\opL p_{\vth}}{p_{\vth}} \ln \frac{p_{\vth}}{p_{\hatvth}} }_{\vth} \right].
\end{aligned}
\]
The relative entropy between two members of a parametric family with parameters $\vth, \hatvth$ is given, to second order in $\hatvth - \vth$, by
\begin{equation} \label{eq:relent_infinitesimal}
D[p_{\vth} \parallel p_{\hatvth}] = \frac{1}{2}(\hatvth - \vth)^{\dagger} G(\vth) (\hatvth - \vth),
\end{equation} 
where $G(\vth) = [G_{kl}(\vth)]$ is the Fisher information matrix of the parametric distribution at parameter value $\vth$, given by
\[ 
G_{kl}(\vth) = \mean{ \frac{\partial \ln p_{\vth}}{\partial \theta_k} \frac{\partial \ln p_{\vth}}{\partial \theta_l}  }_{\vth}.
\]
Using this, we obtain
\begin{align*}
0 &= \nabla_{\hatvth} D[p \parallel p_{\hatvth}] \\
&= G(\vth)(\hatvth - \vth) 
- \Delta t \mean{\nabla_{\hatvth} \ln p_{\hatvth} \frac{\mathcal{L}p_{\vth}}{p_{\vth}} }_{\hatvth}.
\end{align*}
Dividing by $\Delta t$ and taking the limit, we obtain
\[
G(\vth)\dot{\vth} = \mean{(\nabla_{\vth} \ln p_{\vth}) \frac{\opL p_{\vth}}{p_{\vth}}}_{\vth} = \mean{ \opL^{\dagger} \nabla_{\vth} \ln p_{\vth} }_{\vth}
\]
so that the final evolution equation for the parameters $\vth$ reads
\begin{equation} \label{eq:entmatch}
\dot{\vth} = G(\vth)^{-1} \mean{ \opL^{\dagger} \nabla_{\vth} \ln p_{\vth} }_{\vth}.
\end{equation}
We can now see how we can obtain the entropic matching equation \eqref{eq:entmatch} within the variational approach to moment closure: We choose the moment functions $\phi_k = \partial \ln p_{\vth} / \partial \theta_k$.
Then
\[
F_{kl}(\vth)
= \mean{ \phi_k \frac{\partial \ln p_{\vth}}{\partial \theta_l}}_{\vth}
= \mean{ \frac{\partial \ln p_{\vth}}{\partial \theta_k} \frac{\partial \ln p_{\vth}}{\partial \theta_l} }_{\vth} = G_{kl}(\vth)
\]
so that equations \eqref{eq:entmatch} and \eqref{eq:moment_closure} agree.
Note that, in general, we here require the more general form of variational moment closure in which the moment functions depend on the variational parameters $\vth$.
It is also worth noting that, although the relative entropy is non-symmetric in its arguments, 
the final evolution equation for the parameters would be the same had we tried to minimize $D[p_{\vth(t + \Delta t)} \parallel p]$ instead.
This is because infinitesimally, the relative entropy is given by \eqref{eq:relent_infinitesimal}, which is symmetric.

It turns out that \eqref{eq:entmatch} has appeared in the literature previously \cite{brigo1999, brigo2017}, where it was derived by directly defining a projection using the Fisher information metric.
This was done in the context of stochastic filtering equations, which can be considered to be a generalization of a Markovian evolution equation when observations of the stochastic process are included.
Arguably, our derivation could be considered more principled because it does not postulate the use of the Fisher information, but rather starts out with the minimization of the relative entropy, for which as explained above there are strong arguments.
Here we should also mention that more generally, in the context of filtering equations, moment closure approximations are well-known under the name `assumed density filtering' \cite{brigo1999}.

It is also interesting to connect entropic matching to a more advanced information-theoretic approximation for stochastic processes.
Instead of approximating the marginal distributions of the solution of the forward equation \eqref{eq:evolution}, it is possible to approximate the distribution over trajectories of the stochastic process.
This can be done by minimizing the relative entropy of an approximating stochastic process relative to the process of interest over the space of all trajectories.
Such an appoach has been carried out for a number of processes \cite{archambeau2007,opper2008,sutter2016,bravi2017}
and one would expect such an approximation to perform better than entropic matching.
The latter can be understood (for the same reasons already given for variational moment closure) to be a `greedy' algorithm, which at each time point chooses parameters which minimize the relative entropy at an infinitesimally later time point while not taking into account any later time points.
Additionally, entropic matching produces only approximations to single-time marginal distributions and does not provide information about multi-time correlations. This latter deficiency is addressed in Section~\ref{sec:further_app}.
The advantage of entropic matching is that it only requires computations using marginal distributions as given by \eqref{eq:entmatch}.
Variational approximations on the process level, on the other hand, are presumably tractable only for a very small number of processes.

\subsection{Zero-information moment closure} \label{sec:maxent_closure}
It is worthwhile to connect entropic matching with another approach to moment closure motivated by information-theoretic considerations \cite{trebicki1996, smadbeck2013}.
When choosing a distributional ansatz to complement a set of moment functions $\phi_1, \dots, \phi_K$, it seems reasonable to choose the maximum entropy distribution associated with these functions, given by
\begin{equation} \label{eq:maxent}
p_{\vth}(\vx) = \frac{1}{Z(\vth)}\exp\left\{\sum_{k=1}^K \theta_k \phi_k(\vx)\right\} p_0(\vx).
\end{equation}
This is consistent with having the moments $\mean{\vphi(\vx)}_{\vth}$ of the distribution, and no other information available.
The parameters $\vth$ can be computed from the given moments $\mean{\vphi(\vx)}_{\vth}$, and $Z(\vth)$ is a normalization constant. This approach has been termed zero-information (ZI) closure \cite{smadbeck2013}.
Here we have additionally introduced a `background' measure $p_0(\vx)$. This is useful since on an infinite state space, the entropy has in general to be replaced by the relative entropy with respect to some background measure. Equation \eqref{eq:maxent} is then the distribution of minimum relative entropy to $p_0(\vx)$. Note that maximization of entropy has to be replaced by minimization of relative entropy because relative entropy is defined with the inverse sign. 

Entropic matching and ZI moment closure can be seen to be somewhat dual to each other, as is illustrated in Fig.~\ref{fig:relation_diagram}: Entropic matching starts out with a choice of ansatz distribution and supplies a natural choice of moment functions, whereas ZI closure starts out with a choice of moment functions and supplies a natural choice of parametric distribution.
Nevertheless, entropic matching is a generalization of ZI closure, as was first noticed for the case of stochastic differential equations \cite{brigo1999}.
To see this, we check that when entropic matching is applied with a distribution of the form \eqref{eq:maxent}, the result is equivalent to ZI moment closure.
For a distribution of the form \eqref{eq:maxent}, we have $\partial \ln p_{\vth}(\vx)/\partial \theta_k = \phi_k(\vx) - \mean{\phi_k(\vx)}_{\vth}$.
For the matrix $F(\vth)$ from \eqref{eq:matrix_M}, we thus obtain
\begin{align*}
F_{kj}(\vth)
&= \mean{ \phi_k \frac{\partial \ln p_{\vth}}{\partial \theta_j} }_{\vth}
= \mean{ \phi_k (\phi_j - \mean{\phi_j}_{\vth}) }_{\vth} \\
&= \mean{ (\phi_k - \mean{\phi_k}_{\vth}) (\phi_j - \mean{\phi_j}_{\vth}) }_{\vth}\\
&= G_{kj}(\vth),
\end{align*}
i.e. $F(\vth)$ is equal to the Fisher information matrix $G(\vth)$.
Similarly, we have
\[
\mean{ \opL^{\dagger} \phi_k }_{\vth} = \mean{ \opL^{\dagger} (\phi_k - \mean{\phi_k}_{\vth}) }_{\vth}
= \mean{ \opL^{\dagger} \frac{\partial \ln p_{\vth}}{\partial \theta_k}}_{\vth},
\]
so that \eqref{eq:moment_closure} and \eqref{eq:entmatch} agree.

We mention in passing that by using the differential equations \eqref{eq:moment_closure} for the parameters of \eqref{eq:maxent}, the computationally expensive explicit minimization of the relative entropy \cite{smadbeck2013} does not have to be performed \cite{trebicki1996}.

\subsection{Non-uniqueness of the distributional ansatz and minimum relative entropy}
Variational moment closure (and in particular its special case entropic matching) produces an approximation for the distribution of the solution of \eqref{eq:evolution}, as opposed to only approximations for a number of moments.
Ad-hoc moment-closure schemes, on the other hand, are usually considered to only produce approximations for the moments. As explained above, a well-known method to reconstruct full probability distributions from moments is maximum entropy (or more generally minimum relative entropy), and this approach has been used previously in the context of moment equations\cite{trebicki1996,andreychenko2015}.
Here, we briefly point out why such an approach remains meaningful even when a variational moment-closure scheme is employed.

Consider a reaction network with polynomial (e.g. mass-action \eqref{eq:mass_action}) reaction kinetics and assume that we use the standard monomial moment functions \eqref{eq:example_momfuncs} or their high-order analogues.
Choosing a variational ansatz $p_{\vth}(\vx)$, the moment equations \eqref{eq:moment_closure_moments} only ever require a finite number of relations between the moments up to a certain order.
Thus, the distributional ansatz is not uniquely specified by the moment-closure equations derived from it.
This implies that it is not a priori clear how the approximate solution of the CME should be reconstructed from the variational parameters $\vth$ obtained as the solution to \eqref{eq:moment_closure}, and minimum relative entropy provides one possible answer.
The minimum relative entropy distribution again has the form \eqref{eq:maxent} (note however that the parameters $\vth$ governing the minimum relative entropy distribution in \eqref{eq:maxent} are not the same as the parameters of the moment equations \eqref{eq:moment_closure}).
As for ZI moment closure, we mention that an explicit minimization of relative entropy is not necessary \cite{andreychenko2015}, because differential equations for the parameters of the minimum relative entropy distribution can be derived \cite{trebicki1996}. These equations can then be solved simultaneously with the moment equations \eqref{eq:moment_closure}.

%% file: examples.tex
\section{Examples of variational moment closure and entropic matching} \label{sec:entmatch}
In this section, we demonstrate the approximations derived in Section~\ref{sec:theory} on several examples.
Since the standard moment equations using monomial moment functions \eqref{eq:example_momfuncs} are well-known, we here focus on cases where interesting connections to other existing approximations can be established.

\subsection{Product-Poisson entropic matching for the CME} \label{sec:ex:prod_poisson}
A first interesting result is obtained by using a product-Poisson ansatz
\[
p_{\vth}(\vx) = \prod_{n=1}^N{e^{-\theta_n} \frac{\theta_n^{x_n}}{x_n!}}
\]
to approximately solve the CME for mass-action kinetics \eqref{eq:mass_action} via entropic matching. A product-form ansatz might seem to be very restrictive, but it has been shown that a certain class of networks actually has product-form distributions at stationarity \cite{anderson2010, anderson2016}.
Using the backwards evolution operator \eqref{eq:cme_backward} for the CME, we have
\begin{align*}
\opL^{\dagger} \frac{\partial}{\partial \theta_n} \ln p_{\vth}(\vx) 
&= \opL^{\dagger} \frac{\partial}{\partial \theta_n} \sum_{m=1}^N\{x_m \ln \theta_m - \theta_m - \ln x_m! \} \\
&= \frac{1}{\theta_n} \sum_{j=1}^R{h_j(\vx) }\nu_{nj}.
\end{align*}
The factorial moments of a Poisson distribution are given by $\mean{(x)_{s}}_{\theta} = \theta^s$, which allows us to evaluate the right-hand side of \eqref{eq:entmatch}.
The Fisher information matrix of a product Poisson distribution is diagonal, $G_{mn}(\vth) = \delta_{mn} \theta_n^{-1}$.
Combining these results, it turns out that the entropic matching equations are identical with the macroscopic reaction rate equations (RRE)
\begin{equation} \label{eq:poisson_entmatch}
\dot{\vth} = \sum_{j=1}^R{c_j \theta_1^{s_{1j}} \cdots \theta_N^{s_{Nj}} \vnu_j}.
\end{equation}
This implies that the macroscopic rate equations have a meaning even at arbitrary low system sizes.
Product-Poisson entropic matching is an instance of ZI moment closure, where however we have to choose a non-trivial background measure
\[
p_0(\vx) = \frac{e^{-N}}{x_1! \cdots x_N!}.
\]
For some systems, certain molecular species might only exist in zero or one copy, such as a gene which can be either in the active or in the inactive state.
In this case, an analogous result can be obtained by using a Bernoulli distribution instead of a Poisson distribution for the species in question.

\subsection{The finite state projection algorithm for the CME} \label{sec:ex:finite_state}
Variational moment closure encompasses approximations not usually considered as moment closure. For example, taking the ansatz
\begin{equation} \label{eq:finite_state}
p_{\vth}(\vx) = \sum_{\vx' \in \mathbb{X}}\delta_{\vx, \vx'} \theta_{\vx'}
\end{equation}
for some finite subset $\mathbb{X} \subset \natz^N$, and the family of moment functions
\[
\phi_{\vx}(\vx') = \delta_{\vx,\vx'}, \quad \vx \in \mathbb{X},
\]
one recovers the finite state projection algorithm \cite{munsky2006} for the numerical solution of the CME on the finite subset $\mathbb{X}$ of states.
Here we have made use of the fact that within the variational moment closure framework, $p_{\vth}(\vx)$ does not necessarily have to sum to one. This is required because the finite state projection approach `leaks' probability into the part of the state space outside of $\mathbb{X}$.
More generally, the moment functions $\delta_{\vx, \vx'}$ could be replaced by more general basis functions, leading to Galerkin-type approximations \cite{deuflhard2008}.

\subsection{Gaussian entropic matching for the Fokker-Planck equation}
Entropic matching was originally demonstrated \cite{ramalho2013} for the case of a stochastic differential equation (SDE)
\begin{equation} \label{eq:sde}
\df \vx_t = \va(\vx)\df t + S(\vx) \df \mathbf{w}_t
\end{equation}
with a Gaussian distributional ansatz, where however the diffusion matrix $S$ was assumed to not depend on the state $\vx$.
Using our formula, we can extend this result to state-dependent diffusion matrices, beginning with the Fokker-Planck equation (FPE) \eqref{eq:fpe} corresponding to the stochastic differential equation \eqref{eq:sde}.
The details of the computation are provided in Appendix~A.
The resulting evolution equation for mean $\vmu$ and covariance matrix $\Sigma$ of the Gaussian reads
\begin{equation} \label{eq:gaussian_entmatch}
\begin{aligned}
\dot{\vmu} &= \mean{\va(\vx)},\\
\dot{\Sigma} &= \mean{A(\vx)}\Sigma + \Sigma\mean{A(\vx)}^{\dagger} + \mean{B(\vx)},
\end{aligned}
\end{equation}
where $A(\vx)$ is the Jacobian of $\va(\vx)$, $B(\vx) = S(\vx)S(\vx)^{\dagger}$ and the expectation values are taken with respect to the Gaussian approximation.
When $B(\vx) = B$ is state-independent, this agrees with the previous results \cite{ramalho2013}.

It is interesting to note that our result agrees with a process-level approximation \cite{bravi2017} (more precisely, with the single-time marginals) mentioned in Section~\ref{sec:entmatch_theory}.
However, since those results were obtained using a complex form of the relative entropy involving auxiliary variables, it is not clear whether our result would also agree with a proper probabilistic approach using the usual, real relative entropy.

\subsection{Log-normal entropic matching for the CFPE}
For applications in chemical kinetics, a Gaussian ansatz is somewhat unsatisfactory because the ansatz distribution should be restricted to have support on $\pos^N$.
An alternative is to employ a multivariate log-normal distribution.
It turns out that the form of the equations \eqref{eq:gaussian_entmatch} for the parameters $\vmu$ and $\Sigma$ (now parameterizing a log-normal distribution) remain the same. However, the coefficients $\va(\vx)$ and $B(\vx)$ in \eqref{eq:gaussian_entmatch} have to be replaced by
\begin{align*}
\hat{a}_{n}(\vx) &= a_n(e^\vx)e^{-x_n} - \frac{1}{2}e^{-2 x_n} B_{nn}(e^\vx),\\
\hat{B}_{nm}(\vx) &= B_{nm}(e^{\vx})e^{-(x_n+x_m)},
\end{align*}
where for a vector $\vx$, we write $e^{\vx} = (e^{x_1}, \dots, e^{x_N})$.
Note that when we express the entropic matching equations in terms of these modified coefficients, the expectations in \eqref{eq:gaussian_entmatch} have to be computed with respect to a Gaussian distribution with parameters $(\vmu, \Sigma)$, even though these parameters govern a log-normal distribution.
If we employ mass-action kinetics \eqref{eq:mass_action}, the resulting expectations have closed form expressions.

%% file: analysis.tex
\section{Analysis of moment-closure schemes and the Poisson correction} \label{sec:poisson_closures}
General, ad-hoc moment-closure schemes have been shown to exhibit some unphysical properties \cite{schnoerr2014, schnoerr2015}, and attempts to understand these pathologies have so far been unsuccessful.
In this section, we use the variational interpretation of moment closure from Section~\ref{sec:theory} to analyze moment-closure schemes and some their failure modes.

As a first observation, recall that the finite state projection algorithm was shown to be an instance of variational moment closure in Section~\ref{sec:ex:finite_state}.
This simple observation has the important implication that it is not the moment-closure concept as such which leads to unphysical behavior of the resulting approximation.
The finite state projection algorithm will under mild conditions produce an approximation to the exact solution of the CME whose error can be made arbitrary small.
Thus, it is the flexibility of the approximating ansatz distribution (and the choice of moment functions) which determines whether moment closure will produce an acceptable approximation.
The same is true for the RRE, which were also shown to be an instance of ZI closure in Section~\ref{sec:ex:prod_poisson}.

One typical failure mode of ad-hoc moment closures is the divergence of the solutions, especially in the low copy-number regime.
In order to address this problem, we recall that, as explained in Section~\ref{sec:variational}, the distributional ansatz has to be supported on $\natz^N$ or $\pos^N$, depending on whether one considers the CME or the CFPE. This is not difficult to achieve for the CFPE, but it is not obvious for many of the closure schemes introduced previously \cite{lakatos2015} for the CME.
One might conjecture that the divergences observed at low copy numbers might be related to the failure to choose an ansatz distribution with the correct support.
In order to investigate this, we first have to introduce a sufficiently flexible family of distributions with support on $\natz^N$, which we do in the following section.
The approach we present allows one to turn most existing closure methods for the CME into principled closures in the sense of Section~\ref{sec:theory}.
This property makes it particularly convenient to compare ad-hoc closure methods to their principled counterparts.

In this section, we focus on the conventional moment equations using the moment functions \eqref{eq:example_momfuncs} or their higher-order analogues.
We also restrict attention to mass-action kinetics \eqref{eq:mass_action}.

\subsection{Poisson mixtures}
A flexible approach for defining analytically tractable distributions on $\natz^N$ is the mixing of independent distributions on $\natz$ using a distribution on $\pos^N$.
In the context of chemical kinetics, a natural choice for the discrete mixing distributions is the Poisson distribution.
The general form of the distributional assumption that we consider will thus be of the form
\begin{equation} \label{eq:general_poisson_mixture}
p(\vx \mid \vth) = \int_{\pos^N}{\!\!\! \df^N \! \lambda \; p(\vlam \mid \vth) e^{-(\lambda_1 + \cdots + \lambda_N)} \frac{\lambda_1^{x_1} \cdots \lambda_N^{x_N}}{x_1! \cdots x_N!}},
\end{equation}
where $p(\vlam \mid \vth)$ is the mixing distribution governed by parameters $\vth$.
Using Poisson mixtures for the exact or approximate solution of the CME (the Poisson representation) is a well-known approach \cite{gardiner1977}. Here we use it to define a flexible family of distributions for moment closure.
A log-normal mixture of Poisson distributions, used within Eyink's variational framework \cite{eyink1996action}, was proposed previously and motivated using ideas from statistical physics \cite{ohkubo2008}.
It is however important to realize that this approach is simply moment closure using a log-normal-Poisson mixture ansatz.

A situation in which we will be particularly interested in is the case when $p(\vlam \mid \vth)$ is a distribution which has previously been directly employed for moment closure.
This is the case, for example, for the Gamma and multivariate log-normal distributions \cite{lakatos2015}, since these have support on $\pos^N$ and can thus serve as mixing distributions. It is also the case for univariate zero-cumulant (ZC) closure of order two, although this is not immediately obvious.
To perform moment closure, we have to obtain expressions for the expectation values of monomials in $\vx$ for $p(\vx \mid \vth)$.
Using conditional expectations, these are expressed as
\[
    \mean{ x_1^{\alpha_1} \cdots x_N^{\alpha_N} }_{\vth} = \mean{ \pE[x_1^{\alpha_1} \cdots x_N^{\alpha_N} \mid \vlam] }_{p(\vlam \mid \vth)}
\]
where $\pE[\, \cdot \mid \vlam]$ denotes the expectation with respect to a product-Poisson distribution with mean $\vlam$.
For moments of order one to three, for instance, we obtain the equations
\begin{equation} \label{eq:poisson_moments}
\begin{aligned}
\mean{ x_n }_{\vth} &= \mean{ \lambda_n }_{p(\vlam \mid \vth)}, \\
\mean{ x_n x_m }_{\vth} &= \mean{ \lambda_n \lambda_m }_{p(\vlam \mid \vth)} + \delta_{nm} \mean{ x_n }_{\vth},\\
\mean{ x_i x_n x_m }_{\vth} &= \mean{ \lambda_i \lambda_n \lambda_m }_{p(\vlam \mid \vth)} - 2 \delta_{in} \delta_{nm} \mean{ x_i }_{\vth} \\
& + \delta_{mi} \mean{ x_i x_n }_{\vth} + \delta_{in} \mean{ x_n x_m }_{\vth} + \delta_{nm} \mean{ x_m x_i }_{\vth}.
\end{aligned}
\end{equation}
In general, all moments of a product-Poisson distribution are polynomials in the parameters $\vlam$. This implies that if the moment-closure equations using some distribution $p(\vlam \mid \vth)$ for polynomial reaction kinetics can be computed in closed form, then this will also be the case for the Poisson mixture with mixing distribution $p(\vlam \mid \vth)$.
Additionally, we see that the expressions in \eqref{eq:poisson_moments} defining the closure scheme are merely corrected by polynomials in lower-order moments when moving from a closure using $p(\vlam \mid \vth)$ to a Poisson--$p(\vlam \mid \vth)$ mixture closure.
For this reason, we will use the term \emph{Poisson correction} when referring to this case.

One might expect that the difference between a moment-closure scheme defined using a distribution on $\pos^N$ and the corresponding Poisson-corrected closure might be more pronounced for low copy numbers of the chemical species.
Indeed, many moment-closure schemes are known to diverge in this regime.
In the following section, we empirically demonstrate that this problem can be explained by the failure to take into account the discreteness of molecule counts when choosing a moment-closure scheme.

\subsection{A single-species system}
\begin{figure}
\centering
\includegraphics[width=0.70\textwidth]{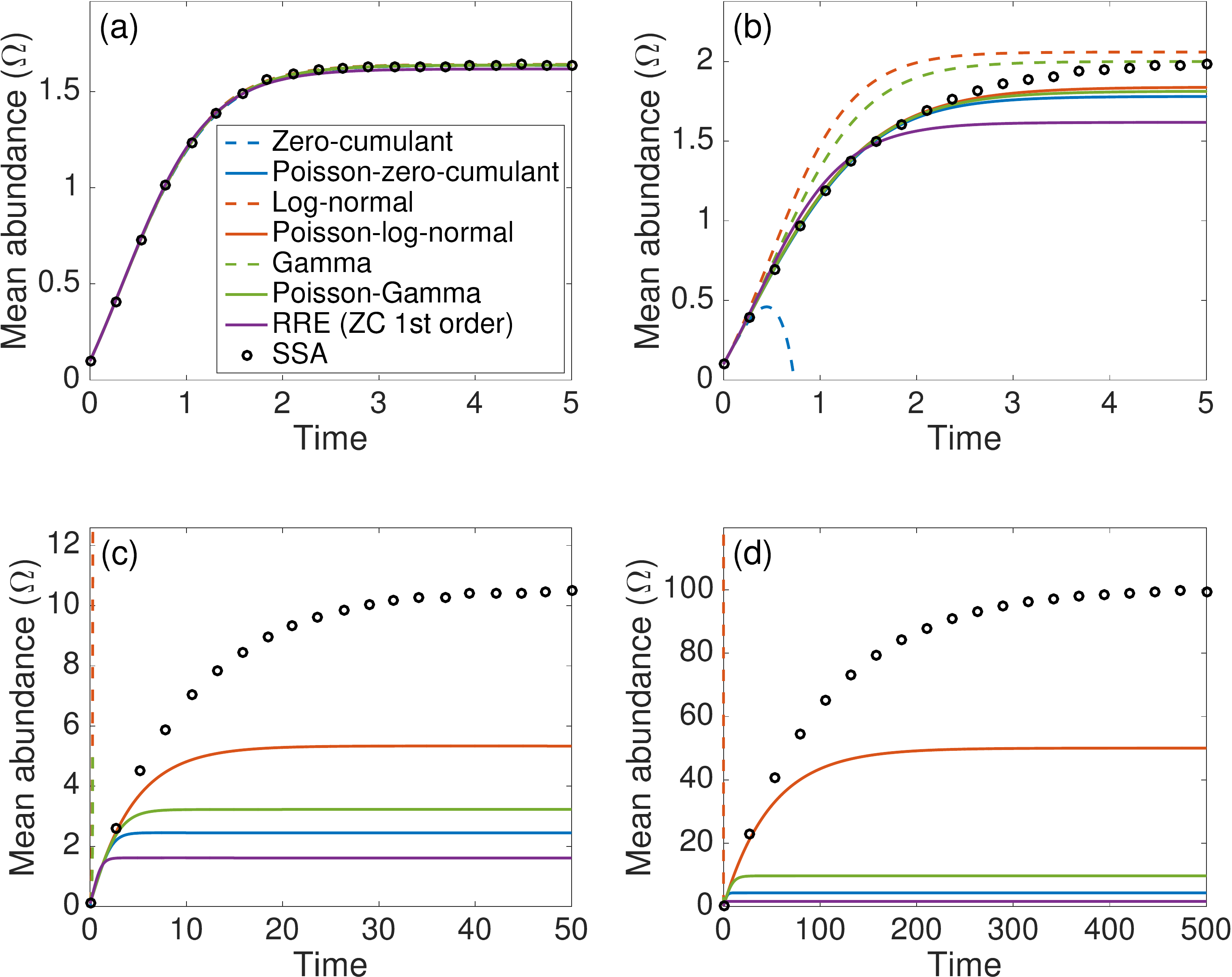}
\caption{Moment closure predictions of mean abundance for \eqref{eq:single_species_eq} at various system sizes for zero-cumulant, log-normal and Gamma closures and their Poisson-corrections. RRE solutions are given for comparison. Abundances are shown in units of $\Omega$. Exact means obtained from 10,000 stochastic simulation algorithm (SSA) realizations. Parameters $a = b = c = 1$. The initial conditions correspond to a Poisson distribution with mean $\Omega/10$. (a) At $\Omega = 10$, all moment closures show negligible error. (b) At $\Omega = 1$, the zero-cumulant closure has diverged. (c) At $\Omega = 10^{-1}$, all closures without Poisson correction have diverged. (d) At $\Omega = 10^{-2}$, approximation quality reduces for all closures except Poisson-log-normal, but the Poisson-corrected closures do not diverge.}
\label{fig:pairann}
\end{figure}

We will investigate the properties of Poisson corrections on the single-species system
\begin{equation} \label{eq:single_species_system}
\emptyset \overset{a}{\rightarrow} \mathcal{X}, \quad \mathcal{X} \overset{b}{\underset{c}{\rightleftharpoons}} 2\mathcal{X}.
\end{equation}
We concentrate on the standard second-order moment equations, which for this system are given by
\begin{equation} \label{eq:single_species_eq}
\begin{aligned} 
\dot{\mu}_1 &= a\Omega + c(\mu_1-\mu_2)/\Omega + b\mu_1, \\
\dot{\mu}_2 &= a\Omega + (2 a\Omega + b - c/\Omega)\mu_1 \\
&\quad + (2 b + 3 c/\Omega)\mu_2 - 2 c \mu_3 /\Omega. 
\end{aligned}
\end{equation}
Here $\mu_1, \mu_2, \mu_3$ are the non-centered moments of order one to three and $\Omega$ is the system size as defined in \eqref{eq:mass_action}.
A moment closure scheme is then given by specifying the third-order moment as a function of the first- and second-order moments, $\mu_3 = V(\mu_1, \mu_2)$.
We focus on three popular ad-hoc closures: The zero-cumulant closure
\begin{equation} \label{eq:closure_zc}
V_{\textrm{ZC}}(\mu_1, \mu_2) = 3\mu_2 \mu_1 - 2 \mu_1^3,
\end{equation} 
the log-normal closure
\begin{equation} \label{eq:closure_ln}
V_{\textrm{LN}}(\mu_1, \mu_2) = (\mu_2/\mu_1)^3
\end{equation}
and the Gamma closure
\begin{equation} \label{eq:closure_gm}
V_{\textrm{G}}(\mu_1, \mu_2) = \mu_2(2\mu_2-\mu_1^2)/\mu_1.
\end{equation}
When applying the Poisson correction, a moment closure function $V(\mu_1, \mu_2)$ is transformed to
\[
\hat{V}(\mu_1, \mu_2) = V(\mu_1, \mu_2-\mu_1) + 3\mu_2 - 2\mu_1.
\]
The result of applying these moment closures on the system is shown in Fig.~\ref{fig:pairann}.
We see that without Poisson correction, the ad-hoc closures diverge for sufficiently small system sizes. This does not happen with the Poisson correction.
We do however see that the quality of the Poisson-corrected closures breaks down at sufficiently small sizes.
For these findings to support our hypothesis, we have to verify that the three ad-hoc closures do not correspond to any distributions defined on $\natz$.
While these closures are defined via distributions on $\pos$ or (in the case of zero-cumulant closure) $\mathbb{R}$, we have to remember that they only specify a relation of the form $\mu_3 = V(\mu_1, \mu_2)$, and there might exist distributions on $\natz$ for which the same relation holds. 
We now show to what extent this is the case.

We are interested in the following questions:
(i) Given a pair $(\mu_1, \mu_2)$, under which conditions does there exist a distribution on $\natz$ with these moments of first and second order?
This defines the domain where a moment closure scheme should ideally be defined.
It also corresponds to the domain of valid initial conditions for moment equations, a fact that has often been neglected in previous studies.
(ii) Assuming that $(\mu_1, \mu_2)$ does correspond to a distribution on $\natz$, and in addition given $\mu_3$, under which conditions does there exist a distribution on $\natz$ with moments of orders one to three given by $(\mu_1, \mu_2, \mu_3)$?
Existence questions of this type have recently been answered \cite{infusino2017}.
Obviously, if $\mu_1 = 0$, the distribution is degenerate and concentrated on $0$.
Thus, in the following, assume $\mu_1 > 0$.
Then it turns out that the well-known condition $\mu_2 \geq \mu_1^2$ is not sufficient for the existence of a distribution with the prescribed moments on $\natz$.
Instead, a necessary and sufficient condition is given by \cite{infusino2017}
\begin{equation} \label{eq:ineq_order_2}
\mu_2 - \mu_1^2 \geq \{\mu_1\}(1-\{\mu_1\}),
\end{equation}
where for any real $y$, we write $\{y\} = y - \lfloor y \rfloor$ with $\lfloor y \rfloor$ the greatest integer smaller or equal to $y$, so that $\{y\}$ is the fractional part of $y$.
Now assume that \eqref{eq:ineq_order_2} holds, and that in addition we are given $\mu_3$.
Then in order for there to exist a distribution on $\natz$ with moments $(\mu_1, \mu_2, \mu_3)$, a necessary and sufficient condition is \cite{infusino2017}
\begin{equation} \label{eq:ineq_order_3}
\frac{\mu_3}{\mu_1} - \left(\frac{\mu_2}{\mu_1}\right)^2 \geq \left\{\frac{\mu_2}{\mu_1}\right\}\left(1-\left\{\frac{\mu_2}{\mu_1}\right\}\right).
\end{equation}
Additionally, if equality holds in \eqref{eq:ineq_order_2}, all higher-order moments are uniquely determined. In particular, $\mu_3$ is then determined by requiring equality in \eqref{eq:ineq_order_3},
\begin{equation} \label{eq:eq_order_3}
\frac{\mu_3}{\mu_1} - \left(\frac{\mu_2}{\mu_1}\right)^2 =  \left\{\frac{\mu_2}{\mu_1}\right\}\left(1-\left\{\frac{\mu_2}{\mu_1}\right\}\right).
\end{equation}
Any principled second-order moment closure $V$ has to satisfy these conditions when $\mu_3$ is replaced by $V(\mu_1, \mu_2)$.
Since we are primarily interested in the low copy-number regime, we note that for $\mu_1 \leq 1$, the domain \eqref{eq:ineq_order_2} of valid pairs $(\mu_1, \mu_2)$ is simply characterized by $\mu_2 \geq \mu_1$.
If we in fact have $\mu_2 = \mu_1$, the resulting equality constraint in \eqref{eq:eq_order_3} reduces to $\mu_3 = \mu_1$.

We proceed to apply these results to the three ad-hoc closure schemes presented above.
\begin{figure*}
\includegraphics[width=1\textwidth]{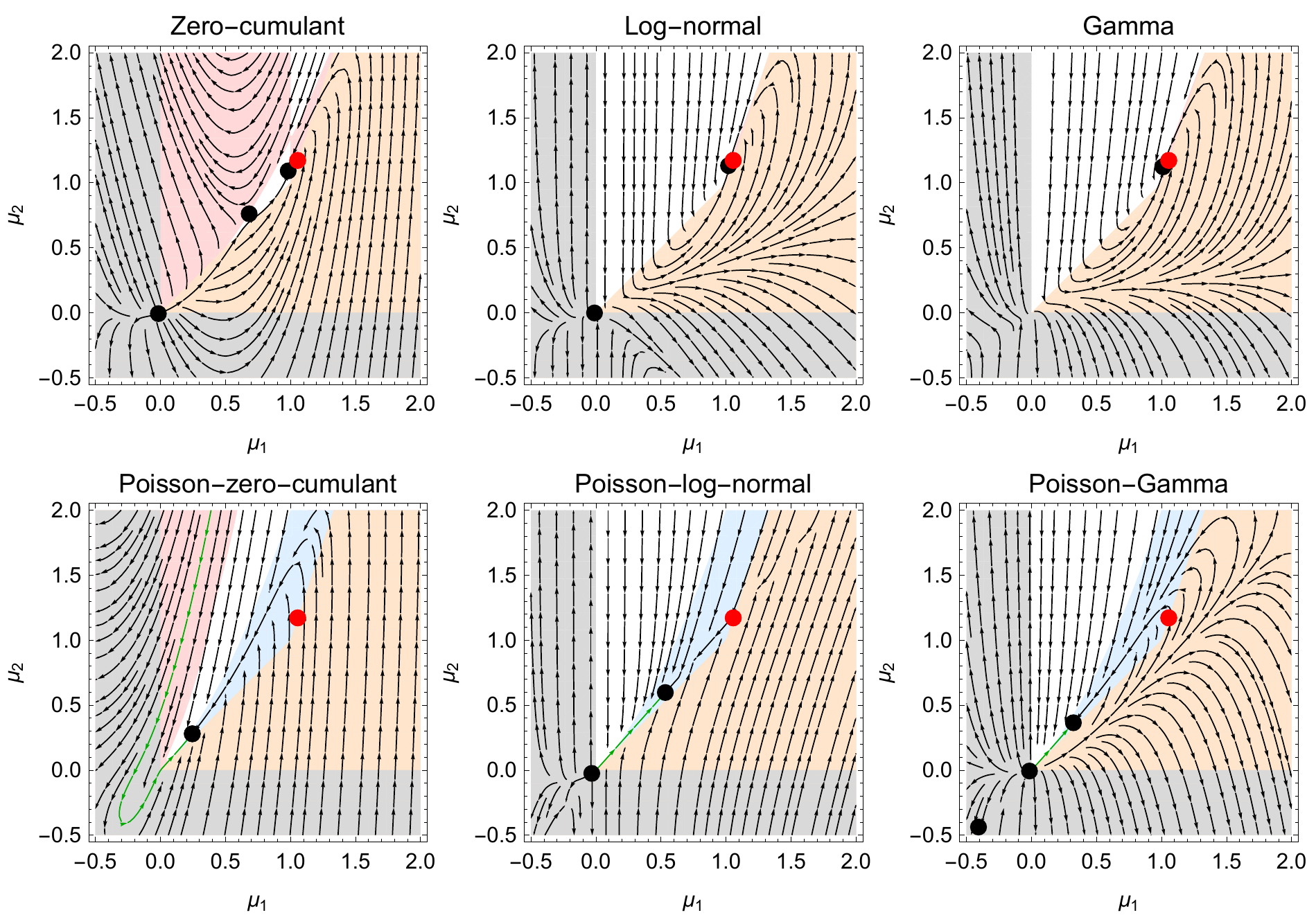}
\caption{Phase-plot for \eqref{eq:single_species_eq} for zero-cumulant, log-normal and Gamma closures (top row) and their Poisson-corrected versions (bottom row).
Red point: Moments of exact stationary distribution. Black point(s): Stationary solution(s) of the moment equations.
Grey area: Region where $\mu_1 < 0$ or $\mu_2 < 0$, shown for better visibility of trajectories. Orange area: Region where \eqref{eq:ineq_order_2} is violated.
Red area: Region where \eqref{eq:ineq_order_2} is valid but \eqref{eq:ineq_order_3} is violated. Blue area: Region where \eqref{eq:ineq_order_2} is valid but the distribution is sub-Poissonian.
Only the white area corresponds to valid initial conditions for the moment equations.
For the Poisson-corrected closures, the green phase curve runs along the domain boundary $\mu_1 = \mu_2$, preventing phase curves from entering the orange region as they do in the ad-hoc closures.
Parameters were $a = b = c = 1$ and $\Omega = 0.1$.}
\label{fig:phaseplots}
\end{figure*}
The regions where \eqref{eq:ineq_order_3} is not satisfied (while \eqref{eq:ineq_order_2} is satisfied) are shown in Fig.~\ref{fig:phaseplots}.
For zero-cumulant closure, the inequality constraint \eqref{eq:ineq_order_3} is not satisfied for most pairs $(\mu_1, \mu_2)$ of region \eqref{eq:ineq_order_2}. For Log-normal and Gamma closures, on the other hand, all pairs $(\mu_1, \mu_2)$ in the region shown do satisfy \eqref{eq:ineq_order_3}.
We would expect that this is reflected in the behavior of these moment closures, and this is verified in Fig.~\ref{fig:phaseplots}, where we show the phase plots for the moment equations \eqref{eq:single_species_eq} in the region of low copy numbers.
Zero-cumulant closure shows pathological behavior for a wide range of initial conditions $(\mu_1, \mu_2)$, whereas log-normal and Gamma closure are well-behaved for a large sub-domain of initial conditions.
However, even for these closures, unphysical behavior does occur.
To understand this, we have to take into account the equality constraint \eqref{eq:eq_order_3} for those pairs $(\mu_1, \mu_2)$ for which equality holds in \eqref{eq:ineq_order_2}. Focusing on the low copy-number regime $\mu_1 \leq 1$, we are concerned with the line $\mu_2 = \mu_1$, and we observe from Fig.~\ref{fig:phaseplots} that it is precisely along this line that unphysical behavior occurs for log-normal and Gamma closures. Indeed, from \eqref{eq:closure_ln} and \eqref{eq:closure_gm}, we find that $V_{\textrm{LN}}(\mu_1, \mu_2) = 1$ and $V_{\textrm{G}}(\mu_1, \mu_2) = (2-\mu_1)\mu_1$ when $\mu_1 = \mu_2$, violating \eqref{eq:eq_order_3}.
In general, the pathologies of the ad-hoc closures seem to correspond quite closely to what one would expect.

We now investigate to what extent these pathologies are removed by applying a Poisson correction.
We first have to determine whether the three ad-hoc closures \eqref{eq:closure_zc}, \eqref{eq:closure_ln} and \eqref{eq:closure_gm} would be compatible with distributions on $\pos$, because they play the role of mixing distributions in \eqref{eq:general_poisson_mixture}.
This is clear for the log-normal and Gamma closures.
For the zero-cumulant closure, this is an instance of the truncated Stieltjes moment problem, and one can check that the relation $\mu_3 = V_{\textrm{ZC}}(\mu_1, \mu_2)$ is compatible with a distribution on $\pos$ as long as $\mu_1^2 \leq \mu_2 \leq 2\mu_1^2$.
This is a relatively small domain, and the behavior visible in Fig.~\ref{fig:phaseplots} is compatible with these findings: Poisson-correction does remove divergences in a relatively small domain.
For log-normal and Gamma closures, we observe from Fig.~\ref{fig:phaseplots} that the divergences close to the line $\mu_1 = \mu_2$ are removed by applying a Poisson correction.
Indeed, one immediately checks that (for $\mu_1 \leq 1$), we have $\hat{V}_{\textrm{LN}}(\mu_1, \mu_2) = \mu_1$ and $\hat{V}_{\textrm{GM}}(\mu_1, \mu_2) = \mu_1$, so that the Poisson-corrected versions do satisfy the equality constraint \eqref{eq:eq_order_3}, as of course they have to.

Our findings suggest that the failure to choose distributions with a support adapted to the CME for moment closure provides a good explanation for the divergences often observed in the low copy-number regime, and that the Poisson correction prevents this from happening.
One further observation from Fig.~\ref{fig:phaseplots} is that the stationary points of the Poisson-corrected closures show a larger error (relative to the exact stationary points) than for the ad-hoc closures.
This is explained by the fact that the system \eqref{eq:single_species_system} at stationarity has a sub-Poissonian distribution, a property which cannot be realized by a mixture of Poisson distributions.
From this, we also see that unphysical behavior of the equations on the one hand, and approximation quality on the other hand are not necessarily correlated.
For completeness, note that the opposite case of distributions with heavier tails than a Poisson distribution can be realized by using an appropriate mixing distribution.

\subsection{Other failure modes}
Apart from divergences, there are other important failure modes of moment closures.
Here we briefly indicate how the variational derivation of moment closures makes these failure modes at least plausible.
The CME, under mild conditions, will have a unique stationary distribution $p(\vx)$ where
\[
\opL p(\vx) = 0.
\]
One would then hope that any approximation should retain this property.
It has however been observed \cite{schnoerr2014, schnoerr2015} that moment closures can posses either multiple stationary states (within a meaningful region of parameters space), or show sustained oscillations and thus fail to converge to a stationary state for certain initial conditions.

However, using our derivation of variational moment closures (or the conceptually analogous derivation of entropic matching), these properties are not surprising and have an intuitive explanation.
Since these approximations can be seen as `greedy' algorithms, choosing the best approximation locally after each infinitesimal time-step, there is no guarantee that the effects which lead to a single stationary distribution can be captured by the approximation.
Thus, for instance, for a system with oscillatory trajectories, the fact that a unique stationary distribution exists is a consequence of the fact that different trajectories do not stay in phase as time progresses.
If an approximation is performed after each infinitesimal time-step, however, this effect is not necessarily captured.

%% file: further_applications.tex
\section{Multi-time joint distributions} \label{sec:further_app}
In this section, we extend variational moment closure and entropic matching to multi-time joint distributions and investigate the conditions under which the resulting approximation is self-consistent with the single-time approximations derived in Section~\ref{sec:theory}.

\subsection{Multi-time moment closure and entropic matching}
Our variational derivation of moment-closure approximations can be extended to an approximation for joint probability distributions at multiple time points.
We here focus on two-time joint distributions, although the same approach extends to higher-order joint distributions.

Fix an initial time-point $t_0$ and consider the two-time joint distribution $p_{t_1,t_0}(\vx^1, \vx^0)$ for which we want to obtain an approximation for $t_1 > t_0$.
In order to derive a variational approximation, we again choose an ansatz distribution $p_{\vth}(\vx^1, \vx^0)$, which now has to depend on two arguments.
For example, any of the Poisson-mixture distributions introduced in Section~\ref{sec:poisson_closures} could be used, where now they have to be defined over a space of dimension $2 N$.
Similarly, we choose a set of moment functions $\vphi(\vx^1, \vx^0) = (\phi_1(\vx^1, \vx^0), \dots, \phi_K(\vx^1, \vx^0))$ which now have to depend on two arguments.
For example, in analogy to the standard moment functions \eqref{eq:example_momfuncs}, we could define the $K = 2 N + N(N+1) + N^2$ moment functions
\begin{equation} \label{eq:example_momfuncs_2}
\begin{aligned}
    \phi^0_{n}(\vx^1, \vx^0) &= x^0_n,  &n&= 1, \dots, N, \\
    \phi^1_{n}(\vx^1, \vx^0) &= x^1_n,  &n&= 1, \dots, N, \\
    \phi^{00}_{nm}(\vx^1, \vx^0) &= x^0_n x^0_m,    &n,m&= 1, \dots, N, n \leq m , \\
    \phi^{10}_{nm}(\vx^1, \vx^0) &= x^1_n x^0_m,  &n,m&= 1, \dots, N, \\
    \phi^{11}_{nm}(\vx^1, \vx^0) &= x^1_n x^1_m,  &n,m&= 1, \dots, N, n \leq m.
\end{aligned}
\end{equation}
These moment functions are again used to define distances between distributions $p(\vx^1, \vx^0)$ and $q(\vx^1, \vx^0)$, 
\begin{equation} \label{eq:error_functions_multitime}
E_k(p, q) = \frac{1}{2}\left[ \mean{\phi_k(\vx^1, \vx^0)}_{p} - \mean{\phi_k(\vx^1, \vx^0)}_{q} \right]^2.
\end{equation}
Here as in Section~\ref{sec:variational} we could again use a more general function $C: \mathbb{R} \rightarrow [0, \infty)$ instead of $(\,\cdot\,)^2/2$ to apply to the difference $\mean{\phi_k}_{p} - \mean{\phi_k}_{q}$ without changing the result.
The derivation of the moment-closure approximation proceeds analogously to the case of single-time marginal distributions.
Assuming again that we have available an approximation $p_{\vth(t_1, t_0)}(\vx^1, \vx^0)$ for some $t_1 \geq t_0$, the evolution of this distribution over a short time-interval $\Delta t$ is given by
\begin{align*}
p(\vx^1, \vx^0) 
&= p_{\vth}(\vx^1, \vx^0) + \Delta t p_{\vth}(\vx^0) \opL_1 p_{\vth}(\vx^1 \mid \vx^0) \\
& \quad + \mathcal{O}(\Delta t^2).
\end{align*}
Here $p_{\vth}(\vx^0)$ and $p_{\vth}(\vx^1 \mid \vx^0)$ are, respectively, the marginal and conditional distributions corresponding to $p_{\vth}(\vx^1, \vx^0)$.
The subscript `1` on the operator $\opL_1$ indicates that it acts only on the argument $\vx^1$, and not on $\vx^0$.
We try to approximate the distribution $p$ by a member of the chosen parametric family by minimizing the distance functions \eqref{eq:error_functions_multitime}.
Writing again $\vth = \vth(t_1, t_0)$, $\hatvth = \vth(t_1+\Delta t, t_0)$, the resulting equations for the minima are
\begin{align*}
0 &= \frac{\partial E_k(p_{\vth} + \Delta t p_{\vth}, p_{\hat{\vth}})}{\partial \hat{\theta}_i} \\
&= \bigg[ \mean{\phi_k(\vx^1, \vx^0)}_{\vth} - \mean{\phi_k(\vx^1, \vx^0)}_{\hat{\vth}}  \\
&\quad + \Delta t \mean{\opL_1^{\dagger}\phi_k(\vx^1, \vx^0)}_{\vth} \bigg] \frac{\partial \mean{\phi_k(\vx^1, \vx^0)}_{\hat{\vth}}}{\partial \hat{\theta}_i},
\end{align*}
where $\opL^{\dagger}_1$ again acts only on the argument $\vx^1$.
Dividing by $\Delta t$ and taking the limit, we obtain as in Section~\ref{sec:variational}
\begin{equation} \label{eq:moment_closure_2}
\dot{\vth} = F(\vth)^{-1} \mean{ \opL_1^{\dagger} \vphi(\vx^1, \vx^0) }_{\vth},
\end{equation}
assuming again that the matrix
\begin{equation*} \label{eq:matrix_M_2}
F_{kj}(\vth) = \mean{ \phi_k(\vx^1, \vx^0) \frac{\partial \ln p_{\vth}(\vx^1, \vx^0)}{\partial \theta_j} }_{\vth}
\end{equation*}
is invertible.
Note that \eqref{eq:moment_closure_2} are equations for the evolution of $\vth = \vth(t_1, t_0)$ in $t_1$, while $t_0$ is fixed.
We can also re-parameterize the equations in terms of the moments $\vmu = \mean{\vphi}_{\vth}$ to obtain
\begin{equation} \label{eq:moment_closure_moments_2}
\dot{\vmu} = \mean{ \opL_1^{\dagger} \vphi(\vx^1, \vx^0) }_{\vmu}.
\end{equation}
We see that the moment-closure equations for two-time joint probability distributions are completely analogous to the case of single-time distributions \eqref{eq:moment_closure}.
The same reasoning can be applied to obtain multi-time joint distributions for entropic matching. The derivation proceeds along similar lines, so that we only state the resulting equation
\begin{equation} \label{eq:entmatch_2}
\dot{\vth} = G(\vth)^{-1} \mean{\opL_1^{\dagger} \nabla_{\vth} \ln p_{\vth}(\vx^1, \vx^0)}_{\vth}.
\end{equation}
For completeness, note that computer implementations for the solution of \eqref{eq:moment_closure} or \eqref{eq:moment_closure_moments} are easily reused for the solution of \eqref{eq:moment_closure_2} or \eqref{eq:moment_closure_moments_2}.
This is because for a reaction system modeled by the CME or the CFPE, the multi-time moment closure equations \eqref{eq:moment_closure_2} and \eqref{eq:entmatch_2} are in fact equivalent to the corresponding single-time equations \eqref{eq:moment_closure} and \eqref{eq:entmatch} applied to an augmented system:
In addition to the species and reactions of the original system \eqref{eq:reaction_system}, we introduce $N$ species $\mathcal{X}_1^0, \dots, \mathcal{X}_N^0$ which do not participate in any reactions. 
Then the single-time equations applied to this $2 N$-dimensional system using an ansatz distribution $p_{\vth}(\vx^1, \vx^0)$ and moment functions $\vphi(\vx^1, \vx^0)$ will be equivalent to the two-time equations.

\begin{figure*}
\centering
\includegraphics[width=1.0\textwidth]{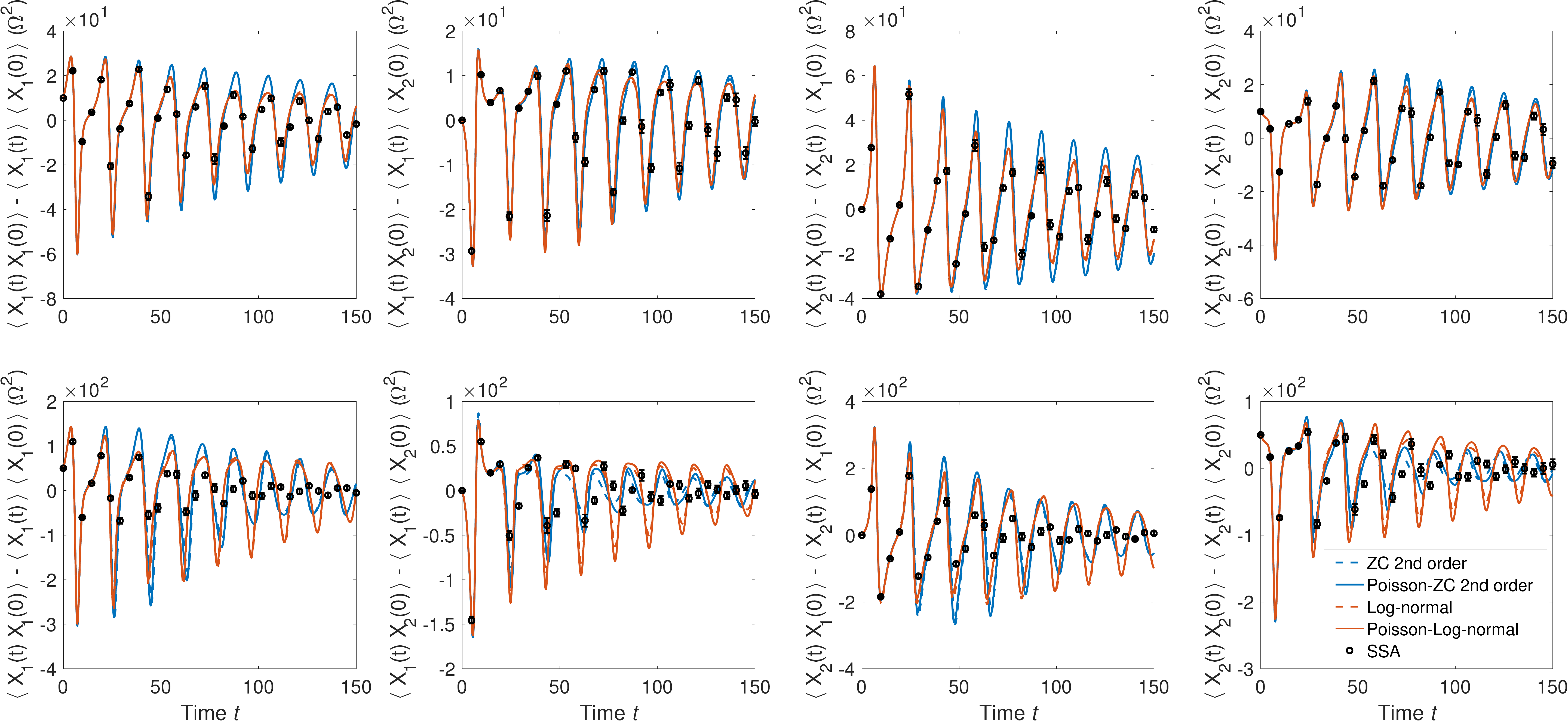}
\caption{Two-time covariance functions $\mean{X_1(t)X_1(0)} - \mean{X_1(t)}\mean{X_1(0)}$, $\mean{X_1(t)X_2(0)}-\mean{X_1(t)}\mean{X_2(0)}$, $\mean{X_2(t)X_1(0)}-\mean{X_2(t)}\mean{X_1(0)}$ and $\mean{X_2(t)X_2(0)}-\mean{X_2(t)}\mean{X_2(0)}$ (corresponding to columns from left to right) as a function of time $t$. System size $\Omega = 5$ (top row) and $\Omega = 1$ (bottom row).
The approximation error at the lower system size is presumably due to the in-adequateness of a second-order closure, regardless of whether a Poisson correction is employed or not.
Parameters values were $c_1 = 1, c_2 = 1, c_3 = 0.5, c_4 = 0.003, c_5 = 0.3$.
Initial conditions were product-Poisson with means $\mean{X_1(0)} = \mean{X_2(0)} = 50 \Omega$.
Black circles correspond to the covariances computed from 100,000 SSA realizations.
Error bars of SSA estimates were computed by dividing the SSA samples into 10 equally sized parts, computing the covariances estimates for each part and plotting $\pm 1$ standard deviation of these estimates.}
\label{fig:joint_lotka}
\end{figure*}

We proceed to demonstrate the two-time moment closure equations numerically.
An interesting case for multi-time correlations are oscillatory systems.
We use the Lotka-Volterra model
\[
\begin{aligned}
\emptyset &\overset{c_1}{\longrightarrow} \mathcal{X}_1,
&\emptyset &\overset{c_1}{\longrightarrow} \mathcal{X}_2, \\
\mathcal{X}_1 &\overset{c_3}{\longrightarrow} 2 \mathcal{X}_1,
&\mathcal{X}_1 + \mathcal{X}_2 &\overset{c_4}{\longrightarrow} 2 \mathcal{X}_2,
& \mathcal{X}_2 &\overset{c_5}{\longrightarrow} \emptyset,
\end{aligned}  
\]
where we added input reactions for each species to prevent explosion and extinction events.
Results for the two-time covariances $\mean{x^1_m x^0_n}_{\vth(t,0)} - \mean{x^1_m}_{\vth(t,0)} \mean{x^0_n}_{\vth(t,0)}$ are shown in Fig.~\ref{fig:joint_lotka}.

\subsection{Self-consistency}
An important question that arises is whether the equations \eqref{eq:moment_closure_2} or \eqref{eq:entmatch_2} produce self-consistent approximations when applied jointly with the single-time approximations \eqref{eq:moment_closure} or \eqref{eq:entmatch}.
The following condition should be satisfied:
Starting from a single-time marginal distribution $p_{t_0}(\vx^0)$ at time $t_0$, we can use \eqref{eq:moment_closure} to produce approximations to the single-time marginals $p_{t_1}(\vx^1)$ for any $t_1 > t_0$. We can also use \eqref{eq:moment_closure_2} to produce approximations to the two-time joint distributions $p_{t_1,t_0}(\vx^1, \vx^0)$, which implies a marginal distribution over $\vx^1$ at time $t_1$.
These two marginal distributions at time $t_1$ should agree.
Of course, this can only hold if the chosen ansatz distributions (which are defined over spaces of different dimensions) and the moment functions are compatible in a sense to be defined below. 
In general, the moment functions $\vphi(\vx^1, \vx^0)$ can be grouped into functions depending on both or on just one of the arguments,
\[
\begin{aligned}
\vphi^1(\vx^1)&, & \vphi^{10}(\vx^1, \vx^0)&, & \vphi^0(\vx^0).
\end{aligned}
\]
Consider a family $p_{\vmu}(\vx^1, \vx^0)$ parameterized in terms of the moments $\vmu = (\vmu^1, \vmu^{10}, \vmu^0)$
\[
\begin{aligned}
\vmu^1 = \mean{\vphi^1}_{\vmu}, \quad \vmu^{10} = \mean{\vphi^{10}}_{\vmu}, \quad
\vmu^0 = \mean{\vphi^0}_{\vmu}.
\end{aligned}
\]
Denote by $p^1_{\vmu}(\vx^1)$ and $p^0_{\vmu}(\vx^0)$ the resulting marginal distributions over $\vx^1$ and $\vx^0$, respectively.
We now assume that $p^1_{\vmu}(\vx^1)$ can be parameterized in terms of $\vmu^1$ only, and similarly $p^0_{\vmu}(\vx^0)$ in terms of $\vmu^0$ only, i.e.
\[
    p^1_{\vmu}(\vx^1) = p^1_{\vmu^1}(\vx^1), \quad p^0_{\vmu}(\vx^0) = p^0_{\vmu^0}(\vx^0).
\]
This is the case, for instance, for the multivariate Gaussian and log-normal distributions, and then also for a log-normal mixture of independent Poissons as introduced in Section~\ref{sec:poisson_closures}.
Then we immediately see that
\begin{equation} \label{eq:marginal_mom_funcs}
\dot{\vmu}^1 = \mean{\opL^{\dagger} \vphi^{1}(\vx^1)}_{\vmu} =  \mean{\opL^{\dagger} \vphi^{1}(\vx^1)}_{\vmu^1}
\end{equation}
because the moment functions $\vphi^1$ do not depend on $\vx^0$, and the marginal distribution $p^1_{\vmu}(\vx^1)$ only depends on $\vmu^1$.
This is a closed equation for the parameters of the marginal distribution $p^1_{\vmu^1}(\vx^1)$.
If, on the other hand, we use the parametric family $p^1_{\vmu^1}(\vx^1)$ and the moment functions $\vphi^1(\vx^1)$ for the single-time moment equations \eqref{eq:moment_closure}, we also obtain \eqref{eq:marginal_mom_funcs}.
Thus, the two-time moment equations using the parametric family $p_{\vmu}(\vx^1, \vx^0)$ are consistent with the single-time moment equations using the parametric family $p_{\vmu^1}(\vx^1)$ as required.
We similarly see that
\[
\dot{\vmu}^0 = \mean{\opL^{\dagger}_1 \vphi^0(\vx^0)}_{\vmu} = \vzero,
\]
because $\vphi^0$ does not depend on $\vx^1$. Thus, the approximation to the marginal distribution at time $t_0$ does not change (i.e. it remains equal to the initially supplied marginal distribution $p^0_{t_0}(\vx^0)$), as one would expect. 

Because of the self-consistency property explained above, our variational approximation is a viable alternative to the process-level variational approximations mentioned in Section~\ref{sec:entmatch_theory}.
Our approximation is tractable for a wide variety of ansatz distributions, whereas closed-form process-level approximations can presumably only be obtained for a very restricted class of ansatz stochastic processes.

%% file: conclusion.tex
\section{Conclusion}
In this article, we derived moment-closure equations as a variational approximation to the solution of the CME and the CFPE, explaining how well-known problems of moment-closure approximations can be either solved or at least understood via our approach.
We generalized entropic matching to arbitrary ansatz distributions and stochastic processes, providing a particularly principled approximation scheme for applications to the kinetics of biomolecular reaction networks. Entropic matching turned out to be a special case of variational moment closure.
Our variational approach allows us to obtain self-consistent approximations for arbitrary multi-time joint distributions. This method does not require computations on the space of trajectories of the underlying stochastic process, greatly increasing the class of ansatz distributions which can be handled in closed form.

The main challenge in the application of these principled approximations is the computation of expectations with respect to the chosen ansatz distribution.
We here considered relatively simple distributional families, for which these expectations had closed-form expressions.
The most interesting extension of our work, from a practical point of view, would be the development of numerical schemes for the efficient treatment of these expectations for more involved distributional families, using, for example, Monte Carlo estimates.
Presumably, in terms of both computational complexity and approximation quality, the resulting approximations would then be somewhat in between full numerical solutions (usually intractable) and the simple closed form approximations considered here and in most other treatments of moment equations.

Having established moment-closure equations as a principled approximation method, we are also in a position to further investigate the properties of moment closures.
We have considered one particular issue in this article, and plan to extend our analysis in future publications.
Another interesting application of our results would be to inference of hidden process states and parameters from observations, where tractable distribution approximations (including multi-time joint distributions) can help in the development of new inference algorithms.

%% file: appendix_gaussian.tex
\section*{Appendix A}
Here we show how to derive the entropic matching equations for an FPE for a Gaussian ansatz distribution.
We consider the FPE with backwards evolution operator \eqref{eq:fpe_backward} corresponding to the (It\^o) stochastic differential equation \eqref{eq:sde}.
The Gaussian ansatz with parameters $\vth = (\vmu, \Sigma)$ is given by
\[
\ln p_{\vth}(\vx) = -\frac{1}{2} \ln \det(2\pi \Sigma) - \frac{1}{2}(\vx-\vmu)^{\dagger}\Sigma^{-1}(\vx-\vmu).
\]
The Fisher information matrix of a Gaussian with respect to the parameters $(\vmu, \Sigma)$ is block diagonal, with one block corresponding to $\vmu$ and given by $\Sigma^{-1}$, and the other block corresponding to $\Sigma$ and given by
\[
\mean{\frac{\partial^2 \ln p_{\vth}}{\partial \Sigma_{ij} \partial \Sigma_{kl}}}
= \frac{1}{2} (\Sigma^{-1})_{ik} (\Sigma^{-1})_{jl}.
\]
We evaluate the right hand side of \eqref{eq:entmatch} and obtain
\[
\begin{aligned}
\mean{ \opL^{\dagger} \frac{\partial \ln p_{\vth}}{\partial \mu_n} }
&= \left[ \Sigma^{-1} \mean{\va(\vx)} \right]_n, \\
\mean{ \opL^{\dagger} \frac{\partial \ln p_{\vth}}{\partial \Sigma_{nm}} }
&= \mean{ \opL^{\dagger} (\vx - \vmu)^{\dagger}\Sigma^{-1} E_{(nm)} \Sigma^{-1} (\vx-\vmu) }\\
&= \frac{1}{2}\mean{\va(\vx)^{\dagger}\Sigma^{-1}E_{(nm)}\Sigma^{-1}(\vx-\vmu)} \\
&\quad + \frac{1}{2}\mean{(\vx-\vmu)^{\dagger}\Sigma^{-1}E_{(nm)}\Sigma^{-1}\va(\vx)} \\
&\quad + \frac{1}{2} (\Sigma^{-1} \mean{B(\vx)} \Sigma^{-1})_{nm} \\
&= \frac{1}{2}(\mean{A(\vx)}\Sigma^{-1})_{nm} + \frac{1}{2}(\Sigma^{-1}\mean{A(\vx)}^{\dagger})_{nm} \\
&\quad + \frac{1}{2} (\Sigma^{-1} \mean{B(\vx)} \Sigma^{-1})_{nm}
\end{aligned}
\]
where $E_{(nm)}$ is a matrix with the entry `1' in the $n$-th row and $m$-th column, and zeros otherwise.
We also used, in the last line, the equation
\[
\mean{u \frac{\partial v}{\partial x_m}} + \mean{v \frac{\partial u}{\partial x_m}}
= \sum_{n=1}^N{\mean{u v (\Sigma^{-1})_{mn}(x_n - \mu_n)}}
\]
for averages of functions $u(\vx), v(\vx)$ with respect to a Gaussian distribution with parameters $(\vmu, \Sigma)$.
Multiplying by the inverse of the Fisher information matrix now yields \eqref{eq:gaussian_entmatch}.

%% file: ms.bbl
\begin{thebibliography}{25}%
\makeatletter
\providecommand \@ifxundefined [1]{%
 \@ifx{#1\undefined}
}%
\providecommand \@ifnum [1]{%
 \ifnum #1\expandafter \@firstoftwo
 \else \expandafter \@secondoftwo
 \fi
}%
\providecommand \@ifx [1]{%
 \ifx #1\expandafter \@firstoftwo
 \else \expandafter \@secondoftwo
 \fi
}%
\providecommand \natexlab [1]{#1}%
\providecommand \enquote  [1]{``#1''}%
\providecommand \bibnamefont  [1]{#1}%
\providecommand \bibfnamefont [1]{#1}%
\providecommand \citenamefont [1]{#1}%
\providecommand \href@noop [0]{\@secondoftwo}%
\providecommand \href [0]{\begingroup \@sanitize@url \@href}%
\providecommand \@href[1]{\@@startlink{#1}\@@href}%
\providecommand \@@href[1]{\endgroup#1\@@endlink}%
\providecommand \@sanitize@url [0]{\catcode `\\12\catcode `\$12\catcode
  `\&12\catcode `\#12\catcode `\^12\catcode `\_12\catcode `\%12\relax}%
\providecommand \@@startlink[1]{}%
\providecommand \@@endlink[0]{}%
\providecommand \url  [0]{\begingroup\@sanitize@url \@url }%
\providecommand \@url [1]{\endgroup\@href {#1}{\urlprefix }}%
\providecommand \urlprefix  [0]{URL }%
\providecommand \Eprint [0]{\href }%
\providecommand \doibase [0]{http://dx.doi.org/}%
\providecommand \selectlanguage [0]{\@gobble}%
\providecommand \bibinfo  [0]{\@secondoftwo}%
\providecommand \bibfield  [0]{\@secondoftwo}%
\providecommand \translation [1]{[#1]}%
\providecommand \BibitemOpen [0]{}%
\providecommand \bibitemStop [0]{}%
\providecommand \bibitemNoStop [0]{.\EOS\space}%
\providecommand \EOS [0]{\spacefactor3000\relax}%
\providecommand \BibitemShut  [1]{\csname bibitem#1\endcsname}%
\let\auto@bib@innerbib\@empty
\bibitem [{\citenamefont {van Kampen}(2007)}]{vankampen1983}%
  \BibitemOpen
  \bibfield  {author} {\bibinfo {author} {\bibfnamefont {N.~G.}\ \bibnamefont
  {van Kampen}},\ }\href@noop {} {\emph {\bibinfo {title} {Stochastic processes
  in physics and chemistry}}}\ (\bibinfo  {publisher} {Elsevier},\ \bibinfo
  {year} {2007})\BibitemShut {NoStop}%
\bibitem [{\citenamefont {Kuehn}(2016)}]{kuehn2016}%
  \BibitemOpen
  \bibfield  {author} {\bibinfo {author} {\bibfnamefont {C.}~\bibnamefont
  {Kuehn}},\ }\bibfield  {title} {\enquote {\bibinfo {title} {Moment closure
  —- a brief review},}\ }in\ \href@noop {} {\emph {\bibinfo {booktitle}
  {Control of self-organizing nonlinear systems}}}\ (\bibinfo  {publisher}
  {Springer},\ \bibinfo {year} {2016})\ pp.\ \bibinfo {pages}
  {253--271}\BibitemShut {NoStop}%
\bibitem [{\citenamefont {Schnoerr}, \citenamefont {Sanguinetti},\ and\
  \citenamefont {Grima}(2014)}]{schnoerr2014}%
  \BibitemOpen
  \bibfield  {author} {\bibinfo {author} {\bibfnamefont {D.}~\bibnamefont
  {Schnoerr}}, \bibinfo {author} {\bibfnamefont {G.}~\bibnamefont
  {Sanguinetti}}, \ and\ \bibinfo {author} {\bibfnamefont {R.}~\bibnamefont
  {Grima}},\ }\bibfield  {title} {\enquote {\bibinfo {title} {Validity
  conditions for moment closure approximations in stochastic chemical
  kinetics},}\ }\href@noop {} {\bibfield  {journal} {\bibinfo  {journal} {The
  Journal of Chemical Physics}\ }\textbf {\bibinfo {volume} {141}},\ \bibinfo
  {pages} {08B616\_1} (\bibinfo {year} {2014})}\BibitemShut {NoStop}%
\bibitem [{\citenamefont {Schnoerr}, \citenamefont {Sanguinetti},\ and\
  \citenamefont {Grima}(2015)}]{schnoerr2015}%
  \BibitemOpen
  \bibfield  {author} {\bibinfo {author} {\bibfnamefont {D.}~\bibnamefont
  {Schnoerr}}, \bibinfo {author} {\bibfnamefont {G.}~\bibnamefont
  {Sanguinetti}}, \ and\ \bibinfo {author} {\bibfnamefont {R.}~\bibnamefont
  {Grima}},\ }\bibfield  {title} {\enquote {\bibinfo {title} {Comparison of
  different moment-closure approximations for stochastic chemical kinetics},}\
  }\href@noop {} {\bibfield  {journal} {\bibinfo  {journal} {The Journal of
  Chemical Physics}\ }\textbf {\bibinfo {volume} {143}},\ \bibinfo {pages}
  {11B610\_1} (\bibinfo {year} {2015})}\BibitemShut {NoStop}%
\bibitem [{\citenamefont {Ramalho}\ \emph {et~al.}(2013)\citenamefont
  {Ramalho}, \citenamefont {Selig}, \citenamefont {Gerland},\ and\
  \citenamefont {En{\ss}lin}}]{ramalho2013}%
  \BibitemOpen
  \bibfield  {author} {\bibinfo {author} {\bibfnamefont {T.}~\bibnamefont
  {Ramalho}}, \bibinfo {author} {\bibfnamefont {M.}~\bibnamefont {Selig}},
  \bibinfo {author} {\bibfnamefont {U.}~\bibnamefont {Gerland}}, \ and\
  \bibinfo {author} {\bibfnamefont {T.~A.}\ \bibnamefont {En{\ss}lin}},\
  }\bibfield  {title} {\enquote {\bibinfo {title} {Simulation of stochastic
  network dynamics via entropic matching},}\ }\href@noop {} {\bibfield
  {journal} {\bibinfo  {journal} {Physical Review E}\ }\textbf {\bibinfo
  {volume} {87}},\ \bibinfo {pages} {022719} (\bibinfo {year}
  {2013})}\BibitemShut {NoStop}%
\bibitem [{\citenamefont {Gillespie}(2000)}]{gillespie2000}%
  \BibitemOpen
  \bibfield  {author} {\bibinfo {author} {\bibfnamefont {D.~T.}\ \bibnamefont
  {Gillespie}},\ }\bibfield  {title} {\enquote {\bibinfo {title} {The chemical
  {L}angevin equation},}\ }\href@noop {} {\bibfield  {journal} {\bibinfo
  {journal} {The Journal of Chemical Physics}\ }\textbf {\bibinfo {volume}
  {113}},\ \bibinfo {pages} {297--306} (\bibinfo {year} {2000})}\BibitemShut
  {NoStop}%
\bibitem [{\citenamefont {Eyink}(1996)}]{eyink1996action}%
  \BibitemOpen
  \bibfield  {author} {\bibinfo {author} {\bibfnamefont {G.~L.}\ \bibnamefont
  {Eyink}},\ }\bibfield  {title} {\enquote {\bibinfo {title} {Action principle
  in nonequilibrium statistical dynamics},}\ }\href@noop {} {\bibfield
  {journal} {\bibinfo  {journal} {Physical Review E}\ }\textbf {\bibinfo
  {volume} {54}},\ \bibinfo {pages} {3419} (\bibinfo {year}
  {1996})}\BibitemShut {NoStop}%
\bibitem [{\citenamefont {Leike}\ and\ \citenamefont
  {En{\ss}lin}(2017)}]{leike2017}%
  \BibitemOpen
  \bibfield  {author} {\bibinfo {author} {\bibfnamefont {R.~H.}\ \bibnamefont
  {Leike}}\ and\ \bibinfo {author} {\bibfnamefont {T.~A.}\ \bibnamefont
  {En{\ss}lin}},\ }\bibfield  {title} {\enquote {\bibinfo {title} {Optimal
  belief approximation},}\ }\href@noop {} {\bibfield  {journal} {\bibinfo
  {journal} {Entropy}\ }\textbf {\bibinfo {volume} {19}},\ \bibinfo {pages}
  {402} (\bibinfo {year} {2017})}\BibitemShut {NoStop}%
\bibitem [{\citenamefont {Brigo}, \citenamefont {Hanzon},\ and\ \citenamefont
  {Le~Gland}(1999)}]{brigo1999}%
  \BibitemOpen
  \bibfield  {author} {\bibinfo {author} {\bibfnamefont {D.}~\bibnamefont
  {Brigo}}, \bibinfo {author} {\bibfnamefont {B.}~\bibnamefont {Hanzon}}, \
  and\ \bibinfo {author} {\bibfnamefont {F.}~\bibnamefont {Le~Gland}},\
  }\bibfield  {title} {\enquote {\bibinfo {title} {Approximate nonlinear
  filtering by projection on exponential manifolds of densities},}\ }\href@noop
  {} {\bibfield  {journal} {\bibinfo  {journal} {Bernoulli}\ }\textbf {\bibinfo
  {volume} {5}},\ \bibinfo {pages} {495--534} (\bibinfo {year}
  {1999})}\BibitemShut {NoStop}%
\bibitem [{\citenamefont {Brigo}\ and\ \citenamefont
  {Pistone}(2017)}]{brigo2017}%
  \BibitemOpen
  \bibfield  {author} {\bibinfo {author} {\bibfnamefont {D.}~\bibnamefont
  {Brigo}}\ and\ \bibinfo {author} {\bibfnamefont {G.}~\bibnamefont
  {Pistone}},\ }\bibfield  {title} {\enquote {\bibinfo {title} {Dimensionality
  reduction for measure valued evolution equations in statistical manifolds},}\
  }in\ \href@noop {} {\emph {\bibinfo {booktitle} {Computational Information
  Geometry}}}\ (\bibinfo  {publisher} {Springer},\ \bibinfo {year} {2017})\
  pp.\ \bibinfo {pages} {217--265}\BibitemShut {NoStop}%
\bibitem [{\citenamefont {Archambeau}\ \emph {et~al.}(2007)\citenamefont
  {Archambeau}, \citenamefont {Cornford}, \citenamefont {Opper},\ and\
  \citenamefont {Shawe-Taylor}}]{archambeau2007}%
  \BibitemOpen
  \bibfield  {author} {\bibinfo {author} {\bibfnamefont {C.}~\bibnamefont
  {Archambeau}}, \bibinfo {author} {\bibfnamefont {D.}~\bibnamefont
  {Cornford}}, \bibinfo {author} {\bibfnamefont {M.}~\bibnamefont {Opper}}, \
  and\ \bibinfo {author} {\bibfnamefont {J.}~\bibnamefont {Shawe-Taylor}},\
  }\bibfield  {title} {\enquote {\bibinfo {title} {Gaussian process
  approximations of stochastic differential equations},}\ }in\ \href@noop {}
  {\emph {\bibinfo {booktitle} {Gaussian Processes in Practice}}}\ (\bibinfo
  {year} {2007})\ pp.\ \bibinfo {pages} {1--16}\BibitemShut {NoStop}%
\bibitem [{\citenamefont {Opper}\ and\ \citenamefont
  {Sanguinetti}(2008)}]{opper2008}%
  \BibitemOpen
  \bibfield  {author} {\bibinfo {author} {\bibfnamefont {M.}~\bibnamefont
  {Opper}}\ and\ \bibinfo {author} {\bibfnamefont {G.}~\bibnamefont
  {Sanguinetti}},\ }\bibfield  {title} {\enquote {\bibinfo {title} {Variational
  inference for {M}arkov jump processes},}\ }in\ \href@noop {} {\emph {\bibinfo
  {booktitle} {Advances in Neural Information Processing Systems}}}\ (\bibinfo
  {year} {2008})\ pp.\ \bibinfo {pages} {1105--1112}\BibitemShut {NoStop}%
\bibitem [{\citenamefont {Sutter}, \citenamefont {Ganguly},\ and\ \citenamefont
  {Koeppl}(2016)}]{sutter2016}%
  \BibitemOpen
  \bibfield  {author} {\bibinfo {author} {\bibfnamefont {T.}~\bibnamefont
  {Sutter}}, \bibinfo {author} {\bibfnamefont {A.}~\bibnamefont {Ganguly}}, \
  and\ \bibinfo {author} {\bibfnamefont {H.}~\bibnamefont {Koeppl}},\
  }\bibfield  {title} {\enquote {\bibinfo {title} {A variational approach to
  path estimation and parameter inference of hidden diffusion processes},}\
  }\href@noop {} {\bibfield  {journal} {\bibinfo  {journal} {Journal of Machine
  Learning Research}\ }\textbf {\bibinfo {volume} {17}},\ \bibinfo {pages}
  {1--37} (\bibinfo {year} {2016})}\BibitemShut {NoStop}%
\bibitem [{\citenamefont {Bravi}\ and\ \citenamefont
  {Sollich}(2017)}]{bravi2017}%
  \BibitemOpen
  \bibfield  {author} {\bibinfo {author} {\bibfnamefont {B.}~\bibnamefont
  {Bravi}}\ and\ \bibinfo {author} {\bibfnamefont {P.}~\bibnamefont
  {Sollich}},\ }\bibfield  {title} {\enquote {\bibinfo {title} {Statistical
  physics approaches to subnetwork dynamics in biochemical systems},}\ }\href
  {http://stacks.iop.org/1478-3975/14/i=4/a=045010} {\bibfield  {journal}
  {\bibinfo  {journal} {Physical Biology}\ }\textbf {\bibinfo {volume} {14}},\
  \bibinfo {pages} {045010} (\bibinfo {year} {2017})}\BibitemShut {NoStop}%
\bibitem [{\citenamefont {Tr\c{e}bicki}\ and\ \citenamefont
  {Sobczyk}(1996)}]{trebicki1996}%
  \BibitemOpen
  \bibfield  {author} {\bibinfo {author} {\bibfnamefont {J.}~\bibnamefont
  {Tr\c{e}bicki}}\ and\ \bibinfo {author} {\bibfnamefont {K.}~\bibnamefont
  {Sobczyk}},\ }\bibfield  {title} {\enquote {\bibinfo {title} {Maximum entropy
  principle and non-stationary distributions of stochastic systems},}\
  }\href@noop {} {\bibfield  {journal} {\bibinfo  {journal} {Probabilistic
  Engineering Mechanics}\ }\textbf {\bibinfo {volume} {11}},\ \bibinfo {pages}
  {169--178} (\bibinfo {year} {1996})}\BibitemShut {NoStop}%
\bibitem [{\citenamefont {Smadbeck}\ and\ \citenamefont
  {Kaznessis}(2013)}]{smadbeck2013}%
  \BibitemOpen
  \bibfield  {author} {\bibinfo {author} {\bibfnamefont {P.}~\bibnamefont
  {Smadbeck}}\ and\ \bibinfo {author} {\bibfnamefont {Y.~N.}\ \bibnamefont
  {Kaznessis}},\ }\bibfield  {title} {\enquote {\bibinfo {title} {A closure
  scheme for chemical master equations},}\ }\href@noop {} {\bibfield  {journal}
  {\bibinfo  {journal} {Proceedings of the National Academy of Sciences}\
  }\textbf {\bibinfo {volume} {110}},\ \bibinfo {pages} {14261--14265}
  (\bibinfo {year} {2013})}\BibitemShut {NoStop}%
\bibitem [{\citenamefont {Andreychenko}, \citenamefont {Mikeev},\ and\
  \citenamefont {Wolf}(2015)}]{andreychenko2015}%
  \BibitemOpen
  \bibfield  {author} {\bibinfo {author} {\bibfnamefont {A.}~\bibnamefont
  {Andreychenko}}, \bibinfo {author} {\bibfnamefont {L.}~\bibnamefont
  {Mikeev}}, \ and\ \bibinfo {author} {\bibfnamefont {V.}~\bibnamefont
  {Wolf}},\ }\bibfield  {title} {\enquote {\bibinfo {title} {Model
  reconstruction for moment-based stochastic chemical kinetics},}\ }\href@noop
  {} {\bibfield  {journal} {\bibinfo  {journal} {ACM Transactions on Modeling
  and Computer Simulation (TOMACS)}\ }\textbf {\bibinfo {volume} {25}},\
  \bibinfo {pages} {12} (\bibinfo {year} {2015})}\BibitemShut {NoStop}%
\bibitem [{\citenamefont {Anderson}, \citenamefont {Craciun},\ and\
  \citenamefont {Kurtz}(2010)}]{anderson2010}%
  \BibitemOpen
  \bibfield  {author} {\bibinfo {author} {\bibfnamefont {D.~F.}\ \bibnamefont
  {Anderson}}, \bibinfo {author} {\bibfnamefont {G.}~\bibnamefont {Craciun}}, \
  and\ \bibinfo {author} {\bibfnamefont {T.~G.}\ \bibnamefont {Kurtz}},\
  }\bibfield  {title} {\enquote {\bibinfo {title} {Product-form stationary
  distributions for deficiency zero chemical reaction networks},}\ }\href@noop
  {} {\bibfield  {journal} {\bibinfo  {journal} {Bulletin of Mathematical
  Biology}\ }\textbf {\bibinfo {volume} {72}},\ \bibinfo {pages} {1947--1970}
  (\bibinfo {year} {2010})}\BibitemShut {NoStop}%
\bibitem [{\citenamefont {Anderson}\ and\ \citenamefont
  {Cotter}(2016)}]{anderson2016}%
  \BibitemOpen
  \bibfield  {author} {\bibinfo {author} {\bibfnamefont {D.~F.}\ \bibnamefont
  {Anderson}}\ and\ \bibinfo {author} {\bibfnamefont {S.~L.}\ \bibnamefont
  {Cotter}},\ }\bibfield  {title} {\enquote {\bibinfo {title} {Product-form
  stationary distributions for deficiency zero networks with non-mass action
  kinetics},}\ }\href@noop {} {\bibfield  {journal} {\bibinfo  {journal}
  {Bulletin of Mathematical Biology}\ }\textbf {\bibinfo {volume} {78}},\
  \bibinfo {pages} {2390--2407} (\bibinfo {year} {2016})}\BibitemShut {NoStop}%
\bibitem [{\citenamefont {Munsky}\ and\ \citenamefont
  {Khammash}(2006)}]{munsky2006}%
  \BibitemOpen
  \bibfield  {author} {\bibinfo {author} {\bibfnamefont {B.}~\bibnamefont
  {Munsky}}\ and\ \bibinfo {author} {\bibfnamefont {M.}~\bibnamefont
  {Khammash}},\ }\bibfield  {title} {\enquote {\bibinfo {title} {The finite
  state projection algorithm for the solution of the chemical master
  equation},}\ }\href@noop {} {\bibfield  {journal} {\bibinfo  {journal} {The
  Journal of Chemical Physics}\ }\textbf {\bibinfo {volume} {124}},\ \bibinfo
  {pages} {044104} (\bibinfo {year} {2006})}\BibitemShut {NoStop}%
\bibitem [{\citenamefont {Deuflhard}\ \emph {et~al.}(2008)\citenamefont
  {Deuflhard}, \citenamefont {Huisinga}, \citenamefont {Jahnke},\ and\
  \citenamefont {Wulkow}}]{deuflhard2008}%
  \BibitemOpen
  \bibfield  {author} {\bibinfo {author} {\bibfnamefont {P.}~\bibnamefont
  {Deuflhard}}, \bibinfo {author} {\bibfnamefont {W.}~\bibnamefont {Huisinga}},
  \bibinfo {author} {\bibfnamefont {T.}~\bibnamefont {Jahnke}}, \ and\ \bibinfo
  {author} {\bibfnamefont {M.}~\bibnamefont {Wulkow}},\ }\bibfield  {title}
  {\enquote {\bibinfo {title} {Adaptive discrete {G}alerkin methods applied to
  the chemical master equation},}\ }\href@noop {} {\bibfield  {journal}
  {\bibinfo  {journal} {SIAM Journal on Scientific Computing}\ }\textbf
  {\bibinfo {volume} {30}},\ \bibinfo {pages} {2990--3011} (\bibinfo {year}
  {2008})}\BibitemShut {NoStop}%
\bibitem [{\citenamefont {Lakatos}\ \emph {et~al.}(2015)\citenamefont
  {Lakatos}, \citenamefont {Ale}, \citenamefont {Kirk},\ and\ \citenamefont
  {Stumpf}}]{lakatos2015}%
  \BibitemOpen
  \bibfield  {author} {\bibinfo {author} {\bibfnamefont {E.}~\bibnamefont
  {Lakatos}}, \bibinfo {author} {\bibfnamefont {A.}~\bibnamefont {Ale}},
  \bibinfo {author} {\bibfnamefont {P.~D.}\ \bibnamefont {Kirk}}, \ and\
  \bibinfo {author} {\bibfnamefont {M.~P.}\ \bibnamefont {Stumpf}},\ }\bibfield
   {title} {\enquote {\bibinfo {title} {Multivariate moment closure techniques
  for stochastic kinetic models},}\ }\href@noop {} {\bibfield  {journal}
  {\bibinfo  {journal} {The Journal of Chemical Physics}\ }\textbf {\bibinfo
  {volume} {143}},\ \bibinfo {pages} {094107} (\bibinfo {year}
  {2015})}\BibitemShut {NoStop}%
\bibitem [{\citenamefont {Gardiner}\ and\ \citenamefont
  {Chaturvedi}(1977)}]{gardiner1977}%
  \BibitemOpen
  \bibfield  {author} {\bibinfo {author} {\bibfnamefont {C.}~\bibnamefont
  {Gardiner}}\ and\ \bibinfo {author} {\bibfnamefont {S.}~\bibnamefont
  {Chaturvedi}},\ }\bibfield  {title} {\enquote {\bibinfo {title} {The
  {P}oisson representation. i. a new technique for chemical master
  equations},}\ }\href@noop {} {\bibfield  {journal} {\bibinfo  {journal}
  {Journal of Statistical Physics}\ }\textbf {\bibinfo {volume} {17}},\
  \bibinfo {pages} {429--468} (\bibinfo {year} {1977})}\BibitemShut {NoStop}%
\bibitem [{\citenamefont {Ohkubo}(2008)}]{ohkubo2008}%
  \BibitemOpen
  \bibfield  {author} {\bibinfo {author} {\bibfnamefont {J.}~\bibnamefont
  {Ohkubo}},\ }\bibfield  {title} {\enquote {\bibinfo {title} {Approximation
  scheme for master equations: Variational approach to multivariate case},}\
  }\href@noop {} {\bibfield  {journal} {\bibinfo  {journal} {The Journal of
  Chemical Physics}\ }\textbf {\bibinfo {volume} {129}},\ \bibinfo {pages}
  {044108} (\bibinfo {year} {2008})}\BibitemShut {NoStop}%
\bibitem [{\citenamefont {Infusino}\ \emph {et~al.}(2017)\citenamefont
  {Infusino}, \citenamefont {Kuna}, \citenamefont {Lebowitz},\ and\
  \citenamefont {Speer}}]{infusino2017}%
  \BibitemOpen
  \bibfield  {author} {\bibinfo {author} {\bibfnamefont {M.}~\bibnamefont
  {Infusino}}, \bibinfo {author} {\bibfnamefont {T.}~\bibnamefont {Kuna}},
  \bibinfo {author} {\bibfnamefont {J.~L.}\ \bibnamefont {Lebowitz}}, \ and\
  \bibinfo {author} {\bibfnamefont {E.~R.}\ \bibnamefont {Speer}},\ }\bibfield
  {title} {\enquote {\bibinfo {title} {The truncated moment problem on
  $\mathbb{N}_0$},}\ }\href@noop {} {\bibfield  {journal} {\bibinfo  {journal}
  {Journal of Mathematical Analysis and Applications}\ }\textbf {\bibinfo
  {volume} {452}},\ \bibinfo {pages} {443--468} (\bibinfo {year}
  {2017})}\BibitemShut {NoStop}%
\end{thebibliography}%
